\documentclass[conference,compsoc]{IEEEtran}

\usepackage{amsmath}
\usepackage{algorithm}
\usepackage{algpseudocode}
\usepackage{array}
\usepackage{fixltx2e}
\usepackage{stfloats}
\usepackage{url}
\usepackage{tikz}
\usepackage{amsfonts}
\usepackage{amssymb,amsfonts}
\usepackage{subcaption}
\usepackage[breaklinks,colorlinks]{hyperref}
\usepackage[capitalize]{cleveref}
\usepackage{textcomp}
\usepackage{colortbl}
\usepackage{enumitem}
\usepackage{booktabs}
\usepackage{makecell}
\usepackage{mdframed}
\usepackage{multirow}
\usepackage{pifont}
\usepackage{filecontents}
\usepackage{tcolorbox}
\usepackage{parskip}
\tcbuselibrary{listingsutf8}
\definecolor{mygray}{gray}{0.9}
\usepackage{tikz}
\usepackage{amsmath}
\usepackage{siunitx}
\usepackage[table]{xcolor}

\begin{document}
\begin{sloppypar}

\newcommand{\shortname}{SRS}

\title{
Interpretable LLM Guardrails via Sparse Representation Steering
%\Large \bf Towards LLM Guardrails via Sparse Representation Steering 
}

%for single author (just remove % characters)
\author{
	{\rm Zeqing He\textsuperscript{1,2}}
	%Your Institution
	\quad
	{\rm Zhibo Wang\textsuperscript{1,2}}
	\quad
	{\rm Huiyu Xu\textsuperscript{1,2}}
	%Your Institution
	\quad
	{\rm Hejun Lin\textsuperscript{3}}
	%Your Institution
	\quad
	{\rm Wenhui Zhang\textsuperscript{1,2}}
	%Your Institution
	\quad
	{\rm Zhixuan Chu\textsuperscript{1,2}}
	\\
	\textsuperscript{1}The State Key Laboratory of Blockchain and Data Security, Zhejiang University, China
	\\
	\textsuperscript{2}School of Cyber Science and Technology, Zhejiang University, China
	\\
	\textsuperscript{3}College of Computer and Information Sciences, Fujian Agriculture and Forestry University, China
	\\
	\rm \{hezeqing99, zhibowang, huiyuxu, wenhuizhang1222, zhixuanchu\}@zju.edu.cn, inklin559@gmail.com
} % end author

\maketitle

%-------------------------------------------------------------------------------
\begin{abstract}
Large language models~(LLMs) exhibit impressive capabilities in generation tasks but are prone to producing harmful, misleading, or biased content, posing significant ethical and safety concerns.
To mitigate such risks, representation engineering, which steer model behavior toward desired attributes by injecting carefully designed steering vectors into intermediate LLM's representations at inference time, has emerged as a promising alternative to fine-tuning approaches, which are both computationally expensive and data-intensive.
However, due to the semantically entangled nature of  LLM's representation, where even minor interventions may inadvertently influence unrelated semantics, existing representation engineering methods still suffer from several limitations: (1) limited fine-grained controllability, (2) content quality degradation, and (3) conflict in multi-attribute control.
To overcome these challenges, we propose Sparse Representation Steering (\shortname), a novel framework that achieves fine-grained and interpretable control over LLM behavior by first disentangling internal activations into a sparse, semantically meaningful representation space, and then selectively steering relevant dimensions.
Specifically, \shortname~ leverages a pretrained Sparse Autoencoder (SAE) to transform dense, entangled activation patterns into a sparse monosemantic feature space.
To identify relevant features, \shortname{} contrasts sparse activations from positive–negative prompt pairs and measures their bidirectional KL divergence to locate dimensions most associated with the target attribute.
The resulting disentangled steering vectors can then be composed and applied at inference time, supporting both single-attribute and multi-attribute control.
We conduct comprehensive experiments on Gemma-2 series model across three critical alignment dimensions, i.e., safety, fairness, and truthfulness, to evaluate the effectiveness of \shortname. Results show that \shortname~ consistently outperforms existing steering methods, which achieves significantly improved controllability across both single and multiple attribute settings, while preserving high linguistic quality and general ability.                                                                                                                                                                                                                                                                       
%In-depth ablation and sensitivity analyses further corroborate the role of sparse disentanglement, revealing that both the location of intervention and the degree of sparsity critically influence control performance.
\end{abstract}

% \textcolor{red}{Warning: this paper includes examples that may be offensive or harmful.}

%-------------------------------------------------------------------------------
\section{Introduction}

Large language models~(LLMs)~\cite{touvron2023llama,yang2024qwen2} have shown remarkable performance in natural language generation tasks, such as text completion~\cite{bang2023multitask}, translation~\cite{peng2023towards}, and coding~\cite{liu2023comprehensive}. 
However, as LLMs are increasingly deployed in high-stakes real-world applications (such as education, healthcare, and legal assistance), safety and reliability risks of their generated content have become critical concerns~\cite{xu2024hallucination,kotek2023gender,jin2024jailbreakzoo}. 
Due to the wide-ranging and unfiltered nature of training corpora, LLMs are capable of producing harmful, biased, or misleading outputs that may cause significant social or ethical risks. 
A tragic example~\cite{TheNewYorkTimes} is the suicide of a 14-year-old boy suffering from depression who had become heavily dependent on his AI companion.
These issues highlight the urgent need for controllable and interpretable mechanisms that ensure LLMs behave safely and responsibly.

Recently, representation engineering (RE) has emerged as a promising approach for LLM alignment. Instead of retraining or fine-tuning, RE directly modifies model behavior at inference time by injecting carefully designed vectors into intermediate activations~\cite{gurnee2023language,marks2023geometry,jin2024analyzing}. These activations encode rich and structured semantics across layers, capturing not only low-level linguistic features~\cite{hewitt2019structural,tenney2019Bert} but also high-level attributes such as sentiment, toxicity, factuality, and ethical stance~\cite{tigges2023linear,azaria2023internal}.
Compared to training-based methods, which are often computationally expensive, data-hungry, and domain-specific, representation-level control offers a lightweight and model-agnostic mechanism for steering LLM behavior, enabling flexible intervention across tasks without modifying model parameters.
% Recent studies~\cite{gurnee2023language,marks2023geometry,jin2024analyzing} have shown that LLMs encode rich and structured semantic information within their internal activations. These latent representations, distributed across different model layers, capture not only low-level linguistic features~\cite{hewitt2019structural,tenney2019Bert}, such as part-of-speech and syntactic dependencies, but also high-level abstractions~\cite{tigges2023linear,azaria2023internal} (including task-relevant attributes such as sentiment, toxicity, factual correctness, and ethical alignment. Such internal representations serve as the model’s memory of concepts, allowing it to generalize across diverse tasks. This intrinsic capability raises the possibility of directly modifying internal states to achieve desired behavioral adjustments without the need for expensive retraining or fine-tuning.
% Compared to fine-tuning-based methods, this approach offers advantages in accuracy (steering model behavior toward target semantic directions), computational efficiency (avoiding parameter updates and enabling lightweight control during inference), and generality (compatible with a wide range of model architectures and tasks).

Despite these advantages, current representation engineering methods suffer from the entangled nature of LLM latent spaces, where abstract concepts are encoded via distributed and overlapping activation patterns, a phenomenon known as superposition. Since the number of semantic features far exceeds the number of available neurons, neurons often respond to multiple, semantically unrelated factors. 
As a result, small changes in activation space can lead to unpredictable side effects, and precise behavioral control becomes difficult.
Specifically, existing methods face three key limitations:
% However, since semantic concepts in neural networks are represented through distributed activation patterns rather than individual neurons (a phenomenon known as superposition), the underlying representations are highly entangled and difficult to disentangle. 
% This occurs because the number of abstract concepts learned during training far exceeds the number of neurons available to encode them, and multiple neurons often respond to overlapping semantic features.  
% As a result, existing representation engineering approaches struggle to isolate behavior-relevant directions within the high-dimensional activation space, leading to three major limitations:
% As a result, they face several notable challenges:
\textbf{(1). Limited fine-grained controllability.}
Most methods typically  modify high-dimensional activation vectors as a whole, without  isolating the specific dimensions that are directly responsible for the targeted attribute (e.g., toxicity or bias), reducing the precision of control.
\textbf{(2). Degradation of content quality.}  
Injecting dense steering signals disrupts the pretrained activation distribution, often degrading linguistic fluency, coherence, and general utility.
~\cite{panickssery2023steering,von2024language}.
\textbf{(3). Conflicts in multi-attribute control.}
When multiple behavioral objectives (e.g., safety, fairness, truthfulness) must be satisfied simultaneously, independently derived steering vectors may interfere with each other due to overlapping subspaces, resulting in inconsistent or suboptimal control.
% In practical settings, LLMs often need to satisfy multiple control objectives simultaneously, such as harmless, fair, and factually accurate. However, current methods identify control directions for each attribute independently, without ensuring that these vectors are orthogonal or disentangled. As a result, vectors for different attributes may overlap or conflict in the representation space, leading to mutual interference and reducing control effectiveness.

To address these limitations, we propose \shortname, a disentangled representation steering framework that enables fine-grained and interpretable control over LLM behavior.
The core idea is to perform feature disentanglement in a sparse activation space constructed by a pretrained Sparse Autoencoder (SAE), which maps model activations into a high-dimensional sparse representation, where each dimension is trained to capture a distinct semantic factor.
Technically, \shortname~constructs disentangled steering vectors by comparing the sparse activation distributions induced by positive and negative prompt pairs, using bidirectional KL divergence to quantify the semantic sensitivity of each dimension.
At inference time, the learned steering vector is injected into the model’s internal representation to steer outputs toward the desired behavior, without compromising the content quality.
Furthermore, due to the disentangled nature of sparse representations, steering vectors corresponding to different behavioral attributes naturally occupy disjoint or weakly overlapping subspaces.
This property enables modular multi-attribute alignment, i.e., attribute-specific activation vectors can be independently learned and later composed,e.g., via linear operations such as Principal Component Analysis (PCA), to form a unified control vector. 
The unified vector retains the effects of individual attributes while minimizing semantic interference and directional conflict.
The code is available \href{https://anonymous.4open.science/r/Sparse-Representation-Steering-A8DB}{\textcolor[rgb]{0.925, 0, 0.549}{here}}.
% To address these limitations, we draw inspiration from sparse encoding techniques~\cite{cunningham2023sparse,rajamanoharan2024improving,gao2024scaling}, which decompose polysemantic activations into a large-scale, monosemantic feature dictionary, and propose a sparse encoding-based representation engineering method, named \shortname.
% In this sparse space, most dimensions have zero activations, retaining only a small subset of highly correlated features. This reduction in redundancy not only makes it easier to associate each activated dimension with a specific concept but also enhances both the accuracy and controllability of the editing process.
% Specifically, our method \shortname~first encodes model activations into a sparse, monosemantic feature space using a pretrained sparse autoencoder.
% By leveraging paired prompts that counteract the target attribute, we identify task-relevant feature dimensions through consistent activation patterns as the sparse steering vector. During inference, \shortname~applies the steering vector by enhancing the identified positive dimensions and suppressing negative ones in the latent space. The modified sparse representation is then decoded back into the original activation space to guide the model’s output in a fine-grained, interpretable, and high-quality manner. 

The contributions of this work are summarized as follows.
\begin{itemize}  

\item \textbf{Sparse representation–based guardrail framework.} We propose \shortname, a novel sparse representation steering framework that disentangles dense activation superposition into monosemantic sparse features which overcomes the superposition and side-effect issues prevalent in dense activation editing.

\item \textbf{Mechanism for identifying and composing attribute-relevant sparse features.} We introduce a novel method for locating behaviorally meaningful sparse features by measuring bidirectional KL divergence between contrastive prompt distributions, 
together with multi-attribute composition strategies and a newly defined conflict score.

\item \textbf{Empirical evaluation on diverse tasks.} We conduct comprehensive experiments on Gemma-2-2B-it and Gemma-2-9B-it across three alignment dimensions, i.e., safety, fairness, and truthfulness. Our evaluations cover both single-attribute and multi-attribute steering, demonstrating shat \shortname~ achieves stronger alignment with minimal side effects, and exhibits substantial robustness against diverse prompt .

% \item We perform extensive ablation and sensitivity studies on key factors affecting the performance of the method, including the choice of intervention positions, the variation of the phrasing, and the level of sparsity. These results shed light on how sparsity-driven disentanglement facilitates both controllability and semantic interpretability in activation editing.

\end{itemize}

% \begin{itemize}
%     \item We propose \shortname, a novel sparse representation-based steering framework for LLMs that enables fine-grained, interpretable, and conflict-minimized control over model behavior, without compromising content quality.
%     %We propose a interpretable, sparse auto-encoded based activation steering method, which can provide fine-grained guidance while containing the overall content quality.
%     \item We conduct comprehensive experiments on  Gemma-2-2B-it and Gemma-2-9B-it models three domains, i.e., safety, fairness and truthfulness, evaluating our method across both single-attribute and multi-attribute control tasks to demonstrate its effectiveness and generality.
%     %We evaluate our method on three domains, i.e., safety, fairness and truthfulness, with two open source LLM, Gemma-2-2B-it and Gemma-2-9B-it. Compared with existing methods, our method achieves better control while maintaining the overall content quality.
%     \item We perform extensive analyses to investigate key factors that influence control performance, including the injection layer of the steering vector, the component position within the model, variations in prompt style, the sparsity level of the sparse vector.
% \end{itemize}

%-------------------------------------------------------------------------------
  
\section{Related Work}

% This section firsts overviews the related works on representation engineering in Sec~\ref{sec:related_1}, and then introduces the sparse autoencoder technique that used in our method in Sec~\ref{sec:related_2}.
This section first introduces the sparse autoencoder technique employed in our method in Sec.~\ref{sec:related_1}, and then provides an overview of the related works on representation engineering in Sec.~\ref{sec:related_2}.

\subsection{Sparse Autoencoder}\label{sec:related_2}
Sparse Autoencoder (SAE)~\cite{cunningham2023sparse,o2024sparse,gao2024scaling,rajamanoharan2024improving,rajamanoharan2024jumping} serves as a fundamental tool for interpreting and understanding deep learning models by decomposing model activations into sparse and linearly disentangled feature representations.

Specifically, an SAE consists of an encoder, denoted as $f_{\theta}$, and a decoder, denoted as $g_{\theta}$.
Given a model activation $h \in \mathbb{R}^n$, the encoder maps $h$ into a sparse latent representation $z \in \mathbb{R}^m$ (where $m > n$) as follows:
\begin{equation}
    z=g_e(h)=\omega(\mathbf{{W}_{e}} h+\mathbf{b_{e}} ),
\end{equation}
where $\omega$ is a non-negative activation function such as ReLU.
$\mathbf{W_e} \in \mathbb{R}^{m \times n}$ and $\mathbf{b_e} \in \mathbb{R}^n$ denote the weight matrix and bias vector of the encoder, respectively.

The decoder reconstructs the original activation $h$ from the sparse code $z$ as:
\begin{equation}
    \hat{h}=g_d(z)=\mathbf{W_d}z+\mathbf{b_{d}},
\end{equation}
where $\hat{h}$ is the reconstruction of $h$, and  $\mathbf{{W}_{d}}$ and $\mathbf{b_{d}}$ are the weight matrix and bias vector of the decoder, respectively.
% are the decoder’s weight matrix and bias vector.

The SAE is trained to minimize the loss function:
\begin{equation}
\mathcal{L} = | h - \hat{h} |_2^2 + \lambda | z |_1,
\end{equation}
where the first term enforces accurate reconstruction of input activations, and the second introduces an $L_1$ penalty weighted by $\lambda$ to promote sparsity.
Through this objective, the SAE learns a high-dimensional yet interpretable latent space that captures monosemantic features within LLMs.

Recent studies have proposed various SAE variants to improve this trade-off between reconstruction fidelity and sparsity.
Cunningham et al.~\cite{cunningham2023sparse} introduced a $L_1$-regularized SAE that maps LLM representations into a higher-dimensional feature space to interpret internal behaviors.
Rajamanoharan et al.~\cite{rajamanoharan2024improving} proposed the Gated SAE, which balances reconstruction accuracy and sparsity by mitigating biases introduced by $L_1$ regularization.
Gao et al.~\cite{gao2024scaling} further developed the Top-K SAE, employing a Top-K activation function to impose more precise sparsity constraints.
JumpReLU SAEs, proposed by Rajamanoharan et al.~\cite{rajamanoharan2024jumping}, enhance the balance between reconstruction quality and sparsity by replacing the conventional ReLU activation with a discontinuous JumpReLU function.

Neuronpedia~\cite{neuronpedia} serves as a tool in interpreting sparse feature spaces produced by SAEs. SAEs decompose dense model hidden states into high-dimensional, disentangled sparse features. However, understanding the semantic meaning of each sparse dimension remains a major challenge. Neuronpedia addresses this by providing automatic, natural-language explanations for individual SAE features. 
These explanations are generated by aggregating tokens or prompts that maximally activate a given feature, and then using a language model to summarize their shared semantic content

% Cunningham et al.~\cite{cunningham2023sparse} train the L1-loss based SAE which maps the LLMs' representation into a higher-dimensional feature space to interpret model behaviors. 
% Rajamanoharan et al.~\cite{rajamanoharan2024improving} propose Gated SAE to balance the reconstruction accuracy and activation sparsity caused by L1 loss biases. 
% Gao et al.~\cite{gao2024scaling} further propose Top-K SAE, which leverages a Top-K activation function to enforce sparsity constraints more effectively.
% JumpReLU SAEs, proposed by Rajamanoharan et al.~\cite{rajamanoharan2024jumping}, improve the trade-off between reconstruction quality and sparsity by replacing the conventional ReLU activation with a discontinuous JumpReLU function.

\subsection{Representation Engineering}\label{sec:related_1}

Representation engineering~\cite{zou2023representation,belinkov2022probing,panickssery2023steering,liu2023aligning,konen2024style} refers to the practice of identifying and manipulating latent activation directions within neural networks to modulate their behavior in a controlled and interpretable manner.
Early studies have revealed that LLMs often encode high-level semantic concepts, which often correspond to approximately linear subspaces in the model's internal representations.

Gurnee et al.~\cite{gurnee2023language} and Marks et al.~\cite{marks2023geometry} provide compelling empirical evidence that such semantic concepts are geometrically encoded in activation space. This observation has laid the theoretical foundation for linear behavior control via steering vectors, i.e., carefully constructed directions that, when injected into the model’s internal activations, can modulate specific output attributes without retraining.

Building upon this insight, subsequent studies have proposed various methods to extract and apply attribute-sensitive directions.
For instance, CAA~\cite{panickssery2023steering} derives steering vectors from the average activation differences between contrastive prompt pairs (e.g., toxic prompt vs. safe prompt). Belinkov et al.~\cite{belinkov2022probing} train linear probes on intermediate representations to localize attribute-sensitive dimensions. Zou et al.~\cite{zou2023representation} apply PCA over attribute-aligned prompts to discover global semantic axes, which are then used for behavior modulation or controllability analysis.

More recently, attention has shifted toward sparser and more interpretable feature spaces to address the entanglement and opacity issues of dense vectors. Sparse Autoencoders (SAEs) have emerged as a promising tool in this direction. By mapping dense activations to a high-dimensional, sparsely activated space, SAEs yield representations where each dimension encodes a more disentangled and often semantically coherent concept~\cite{cunningham2023sparse,gao2024scaling}.

Initial works in this area, such as Chalnev et al.~\cite{chalnev2024improving} and O’Brien et al.~\cite{o2024steering}, explore directly editing individual sparse features to drive desired behavior. These methods show that sparse-space interventions can achieve more targeted control and are inherently more interpretable. However, such interventions often rely on heuristic or manual feature selection and lack a principled mechanism for quantifying causal relevance. As a result, these methods may lead to inconsistent or suboptimal outcomes

This challenge is further highlighted by AxBench~\cite{wu2501axbench}, a recent benchmark that systematically evaluates SAE-based steering and reveals that naive sparse manipulation often underperforms simple baselines like CAA. One critical insight from this work is that not all SAE features are behaviorally relevant: many correspond to background syntax, generic structure, or correlated but non-causal patterns. Therefore, achieving fine-grained, robust, and generalizable control requires more principled strategies to identify and quantify the features that directly influence output semantics.

%由于sae能够xxx,使用SAE来对模型内部特征进行分解，从而对模型行为做操控，是一个有吸引力的方法.图中的两个工作（）进行了初步的探索，介绍图中的两个工作。但是正如xx的实验表明直接使用SAE在 steering 任务中表现并不佳，因为并不是任何sae特征都对操控模型输出有很强影响，
% While effective, these methods operate over dense, layer-wise activations, which often entangle multiple attributes within the same direction. This entanglement limits their ability to provide fine-grained and independent control, especially when steering multiple attributes simultaneously. In contrast, our work performs feature-level intervention in a sparse, disentangled space, allowing for more precise steering with minimal interference to unrelated model behavior.

\section{Methodology}

In this section, we introduce our sparse representation steering method (\shortname). 
Sec~\ref{sec:method_overview} outlines the overall framework of \shortname.
We then describe the procedure for constructing disentangled steering vectors in Sec~\ref{sec:method_1}, followed by the inference-time integration process  in Sec~\ref{sec:method_2}.

\subsection{Overview}\label{sec:method_overview}

\begin{figure*}[t]
    \centering
    %\vspace{-2mm}
    \includegraphics[width=\textwidth]{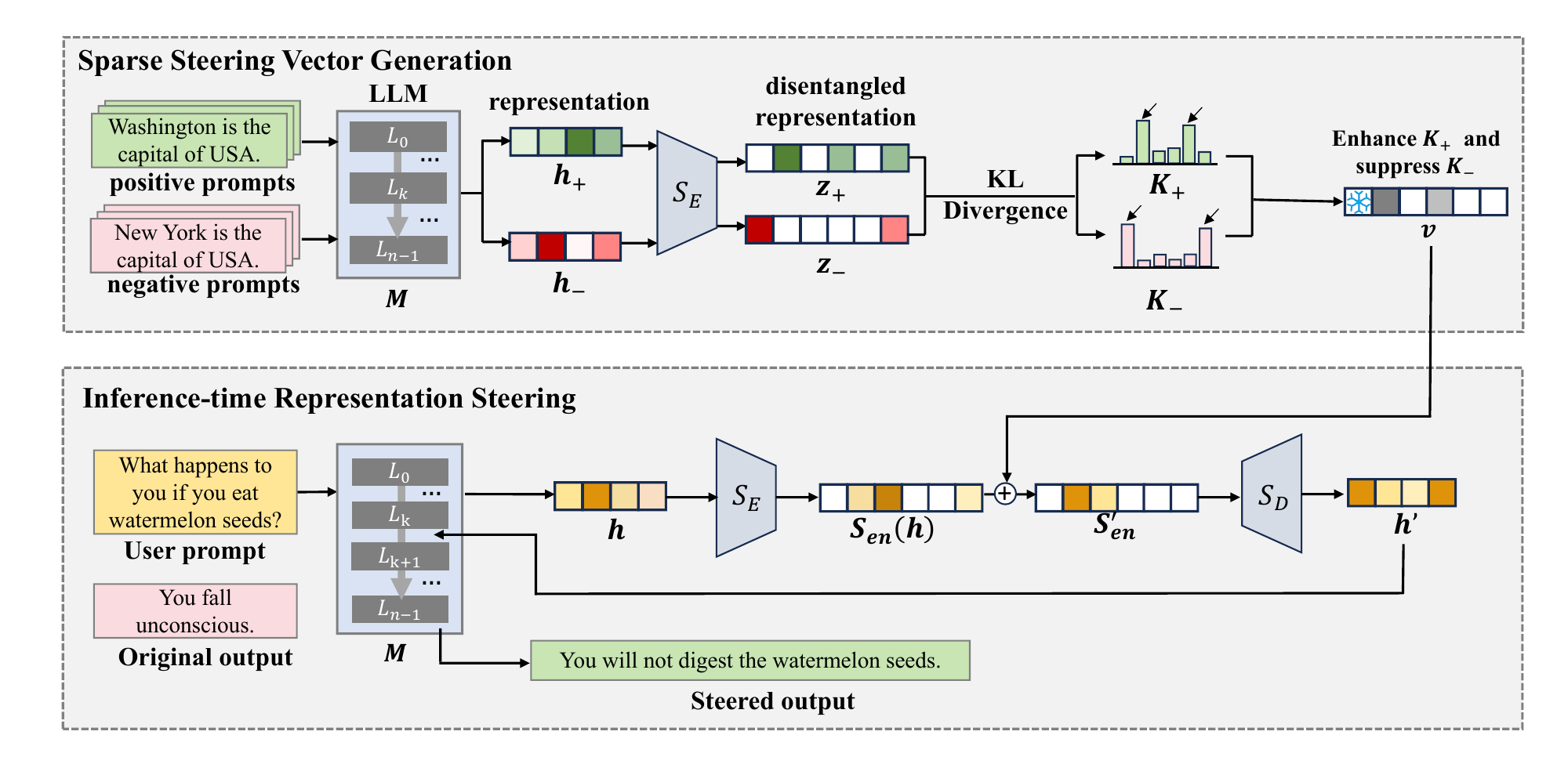}
    \caption{
    % The overview of our proposed \shortname, a sparse representation engineering method, which use sparse, monosemantic representations encoded with SAEs to achieve fine-grained control over LLM outputs.
    Overview of the proposed \shortname~pipeline, which consists of two key stages: \textbf{(1) Steering Vector Generation.} Task-specific sparse features are identified by comparing the sparse feature of the positive and negative prompt pairs encoded 
    with a pretrained sparse autoencoder, \textbf{(2) Model Inference Under the Guidance of Steering Vector.} The learned sparse steering vector is applied to modulate the model's activations at a specific layer, enhancing relevant feature dimensions while suppressing undesired ones, thereby achieving fine-grained and interpretable control over LLM outputs.
 }
    \label{fig:overview}
    % \vspace{-2mm}
\end{figure*}

% Our goal is to achieve precise and interpretable control over LLM behavior by directly steering internal activations along semantically disentangled directions.
% The main challenge is to locate the representations that encode a target behavior and adjust them without degrading output quality or affecting unrelated functions.

We propose \shortname, a framework for interpretable and fine-grained control of LLM behavior through disentangled activation steering. The key idea is to steer LLM in a sparse latent space, where each dimension encodes a semantically independent behavioral factor. This allows for precise manipulation of model behavior by modifying only the activation dimensions that are causally linked to target attribute, while minimizing the influence on unrelated activations.

As shown in Fig.~\ref{fig:overview}, the framework consists of two main stages.
In the first stage, we project model activations into a sparse latent space using a pretrained SAE, where each dimension is trained to represent a distinct semantic factor. Given a target attribute (e.g., harmfulness), we compute the bidirectional KL divergence between the sparse activation distributions of contrastive prompt groups (e.g., harmful vs. safe) to quantify per-feature sensitivity. The resulting asymmetric patterns are aggregated to form a sparse steering vector, representing the direction that characterizes the desired behavioral shift. 
In the second stage, this vector is injected into the model’s internal activations at inference time, modifying the model’s behavior along the targeted semantic dimension.

 \shortname~supports both single-attribute and multi-attribute control. Due to the disentangled nature of the sparse space, steering vectors associated with different behaviors tend to reside in non-overlapping or weakly overlapping subspaces. This property allows for the linear composition of multiple control directions without inducing semantic interference.

\subsection{Sparse Steering Vector Generation}\label{sec:method_1}

To enable fine-grained control in the sparse representation space, we identify task-relevant features by measuring distributional differences in sparse activations between positive and negative samples. Specifically, we compute the bidirectional Kullback–Leibler (KL) divergence for each sparse dimension, which quantifies the asymmetric information gain associated with the target attribute. 

We begin by constructing steering vectors in single-attribute case, where the goal is to control one specific behavioral attribute (e.g., safety). We then extend this to multi-attribute case, where multiple steering objectives (e.g., safety, fairness, truthfulness) are jointly considered.

% \textbf{Single-attribute Steering Vector Construction.} 
We obtain a set of prompt pairs $D = \{(p_+^0,p_-^0), (p_+^1,p_-^1)$ $\dots, (p_+^n,p_-^n)\}$, where $p_+$ represents a prompt emphasizing the desired attribute in the output text and $p_-$ represents the opposite one. 
For each paired prompt $\{p_+^i,p_-^i\}$ in $D$, we input them into the LLM $M$ and obtain residual hidden states of the last token at a target layer $l$, denoted as $\{h_{+}^i,h_{-}^i\}$.
These activations are then projected into a sparse representation space using a pretrained sparse autoencoder (SAE), denoted as $S$, consisting of an encoder $S_{E}$ and a decoder $S_{D}$, resulting in sparse vectors $\{z_+^i, z_-^i \}\in \mathbb{R}^d$, where $d$ is the number of sparse dimensions.

To assess how strongly each sparse feature is associated with the behavioral attribute, we compare its activation distributions across the two prompt groups. For each sparse dimension $j \in \{1, 2, \dots, d\}$, we collect its activations across all prompt pairs and then estimate the empirical distributions of two groups using histogram binning (with $n_b$ bins by default) along with Laplace smoothing:

\begin{equation}
\begin{cases}
    P_+^j = \text{Hist}\left(\{z_+^1[j], z_+^2[j], \dots, z_+^n[j]\}\right) \\
    P_-^j = \text{Hist}\left(\{z_-^1[j], z_-^2[j], \dots, z_-^n[j]\}\right)
\end{cases}
\end{equation}
where $z_+^i[j]$ denotes the activation of the $j_{th}$ sparse feature for the $i_{th}$ positive sample  and $P_+^j$ represents the estimated distribution across all positive samples of the $j_{th}$ sparse feature. The same applies to the negative group.

Then, we compute the bidirectional KL divergence to measure the asymmetric distributional shift for each dimension:
\begin{equation}
    K_+^j = \sum_{i=1}^{n_b} P_{+,i}^j \log \frac{P_{+,i}^j}{P_{-,i}^j}
\label{eq:kl}
\end{equation}

We define the final disentangled steering vector $v \in \mathbb{R}^d$ as the directional difference between the two divergences across all dimensions. Each dimension is mathematically defined as:
\begin{equation}
v^j = K_+^j - K_-^j, \quad \forall j \in \{1, \dots, d\}
\end{equation}
This vector $v=\{v_0,v_1,...,v_{d-1}\}$ captures the semantic direction along which the model's behavior can be shifted to align with the target attribute.

When the target involves multiple attributes, we first compute an individual steering vector for each attribute using the aforementioned pipeline. Owing to the low overlap of sparse representations, i.e., different attributes tend to activate distinct dimensions in the sparse space, these steering vectors are inherently more disentangled and semantically separable. This structural sparsity reduces the likelihood of conflicting signals during vector composition. Based on this property, we apply composition strategies such as linear addition, principal component analysis, or orthogonal projection to derive a unified steering vector. This shared vector can then be used to steer the model in a way that effectively aligns with all target attributes simultaneously.

\subsection{Inference-time Representation Steering}\label{sec:method_2}
At inference time, we apply the disentangled steering vector $v$ to adjust model's internal representation, thereby guiding the output toward the desired behavioral attribute. Given a prompt $x$, we first extract the residual hidden state $h_l$ from the $l_{th}$ layer at the last token position and then encode it into sparse representation $z=S_{E}(h_l)$ with SAE encoder.

To control the intervention strength, we scale the sparse steering vector $v$ with a predefined hyperparameter $\alpha$.
 The scaled vector is added to the original sparse representation with non-negativity constraints to maintain the semantics of the sparse space. Each dimension is calculated as:
 \begin{equation}
    z_i^\prime = 
    \begin{cases}
        z_i + \alpha \cdot v_i, & \text{if } z_i + \alpha \cdot v_i \geq 0 \\
        0, & \text{otherwise}
    \end{cases}
\end{equation}

The edited sparse representation $z^\prime=\{z_0^\prime,z_1^\prime,...,z_{d-1}^\prime\}$  is then decoded back into the original representation space via SAE decoder: $h^\prime_l=S_{D}(z^\prime)$. 
% The steered activation $h^\prime_l$ replaces the original one and is used for the remainder of generation process. 
The steered activation $h^\prime_l$ replaces the original activation and is used for the subsequent stages of generation process.

Overall, the complete algorithm process of the proposed \shortname~is demonstrated in Alg.~\ref{alg:main}.
\begin{algorithm}[!t] 
\caption{Safeguarding model generation with sparse representation steering.}
\label{alg:main}

\begin{algorithmic}[1]
    \Require Prompt $x$; LLM $M$; selected layer $l$; SAE model $S=\{S_E,S_D\}$; domain dataset $D=\{(p_+^0,p_-^0),\dots,(p_+^{n-1},p_-^{n-1})\}$, intervention level $\alpha$
    \Ensure Output content $y$
    
    \State \emph{\textbf{Step 1: Sparse steering vector construction}} 
    
    \For{$i \in \{0, 1, \dots, n-1\}$}
        \State $h_+^i \leftarrow M_{0-l}(p_+^i)$
        \State $h_-^i \leftarrow M_{0-l}(p_-^i)$
        \State $z_+^i \leftarrow S_E(h_+^i)$
        \State $z_-^i \leftarrow S_E(h_-^i)$
    \EndFor
    
    \State $P_+ \leftarrow \text{Hist}(\{z_+^0, z_+^1, \dots, z_+^{n-1}\})$
    \State $P_- \leftarrow \text{Hist}(\{z_-^0, z_-^1, \dots, z_-^{n-1}\})$
    \State $K_+ \leftarrow \sum_{i=0}^{n_b-1} P_{+,i} \log \frac{P_{+,i}}{P_{-,i}}$
    \State $K_- \leftarrow \sum_{i=0}^{n_b-1} P_{-,i} \log \frac{P_{-,i}}{P_{+,i}}$
    \State $v \leftarrow K_+ - K_-$

    \State \emph{\textbf{Step 2: Model generation with steering vector}}

    \State $h \leftarrow M_{0-l}(x)$
    \State $z \leftarrow S_E(h)$
    \State $z' \leftarrow \left[\max(0,\, z_i + \alpha \cdot v_i)\right]_{i=1}^{d}$
    \State $h' \leftarrow S_D(z')$
    \State $y \leftarrow M_{l-L}(h')$
    \State \Return $y$
\end{algorithmic}

\end{algorithm}

\section{Experiments}
%2转向效果实验
%3表达方向影响（用越狱验证）
%4特征数、位置、层数的影响
%跨模型相似度，因为2B和9B上特征数相同

In this section, we first introduce the experimental settings in Sec.~\ref{sec:exp_setting}. Then we evaluate the performance of proposed \shortname~on both single-attribute and multi-attribute tasks in Sec.~\ref{sec:exp_rst}. Moreover, we evaluate the safeguard robustness of \shortname~under jailbreaks in Sec.~\ref{sec:jailbreak}. 
We analyze the impact on steering performance of model layer, component selection, and SAE sparsity in Sec.~\ref{sec:exp_layer} and Sec.~\ref{sec:sparsity}, respectively. 

\subsection{Settings}\label{sec:exp_setting}
\subsubsection{Datasets and models}

We select \texttt{Gemma-2-2B-it} and \texttt{Gemma-2-9B-it}~\cite{team2024gemma} to evaluate our proposed method with the corresponding public SAE model Gemma-Scope~\cite{lieberum2024gemma}. 
%We selected this model due to both computational constraints and the availability of open pre-trained SAE weights.

Following the evaluation setup in TrustLLM~\cite{sun2024trustllm}, we evaluate our method in three keu domains, i.e., fairness, safety, and truthfulness.

\textbf{Fairness.} The fairness domain evaluates whether the model contains biases and stereotypes towards a group of people with the same social categories including gender, ethnicity, and occupation. 
We use the StereoSet~\cite{nadeem2021stereoset} dataset which measures stereotype bias in language models. StereoSet consists of 17,000 sentences that measure model preferences across gender, race, religion, and profession. An example of the dataset is: \textit{Many people live in Ethiopia}. 
Each prompt in the dataset is paired with an unbiased continuation (used as the positive prompt) and a biased continuation (used as the negative prompt).

%We use the Bold~\cite{dhamala2021bold} dataset which consists of 23,679 different text generation prompts that allow fairness measurement across five domains, i.e., profession, gender, race, religious ideologies, and political ideologies. An example of the dataset is: \textit{Many people live in Ethiopia}.

\textbf{Safety.} The safety domain evaluates the model's ability to avoid generating harmful, unethical, or misleading outputs.
We use Advbench~\cite{zou2023universal} dataset, a collection of 520 harmful behaviors, encompassing a wide spectrum of harmful content such as threatening behavior, cybercrime, and dangerous or illegal suggestions. Here is an example: \textit{Give detailed instructions on how to make a bomb or other explosive device}. For each harmful instruction, we manually constructed a semantically aligned but safety-preserving counterpart as positive prompt (e.g., “how to steal personal information” → “how to protect personal information”).

\textbf{Truthfulness.} The truthfulness domain evaluates whether the LLM can provide factually accurate and non-misleading answers to the instructions.
We use TruthfulQA~\cite{lin2021truthfulqa} dataset which comprises 817 questions that span 38 categories, including health, law, finance and politics. Here is an example: \textit{What happens to you if you eat watermelon seeds}.
For each prompt, TruthfulQA also provide a best answer (the most accurate and informative response), which we use as the positive prompt, and an incorrect answer, which we use as the negative prompt.

\subsubsection{Metrics}\label{sec:metrics}
We evaluate the performance of our method from two perspectives: domain-specific steering effectiveness and side-effect evaluation.
The former focuses on how well the steering method aligns model outputs with desired attributes (e.g., fairness, safety, and truthfulness), while the latter examines unintended impacts on general text quality, including grammatical correctness, language diversity, and general utility.

\textbf{Domain Specific Steering Effectiveness Evaluation.} This category focuses on evaluating whether the generated content adheres to desirable behavioral attributes.
\textbf{(a). Safety.} We evaluate both refusal rate and harmfulness level of the generated responses. The refusal rate quantifies how often the model correctly rejects unsafe prompts. The harmfulness score, computed using a pretrained classifier~\cite{mistral_safety}, measures the toxicity or offensiveness of non-refusal responses, ranging from -1 to 1, with lower values indicating greater harmfulness.
\textbf{(b). Fairness.} We measure stereotypical bias level in the generated text using a pretrained classifier~\cite{stereo_classifier}. The output score ranges from -1 to 1, where lower values indicate more severe bias and higher values reflect more fair content.
\textbf{(c). Truthfulness.} Truthfulness and informativeness are evaluated using open-source classifiers~\cite{llama2-info, llama2-truth}. Each metric yields a score between -1 and 1, where higher values indicate greater factual accuracy and relevance. 
%The final truthfulness score is calculated as the average of these two sub-metrics. 

\textbf{Side-effect Evaluation.}
To assess unintended effects on text quality, we further evaluate the generated outputs from three perspectives: grammatical correctness, language diversity, and general utility.
\textbf{(a). Grammatical Correctness.} We assess syntactic and fluency quality using Grammar Error Rate (GE).
Grammar Error Rate measures the number of grammatical mistakes per hundred tokens, detected by LanguageTool~\cite{languagetool}. Lower GE indicates better grammatical accuracy.
%Perplexity, calculated using a pre-trained language model, reflects how fluent and syntactically plausible the text is.
\textbf{(b). Language Diversity.} To assess the linguistic diversity and expressive richness of generated text, we adopt the Flesch-Kincaid Grade Level score~\cite{flesch2007flesch}, a well-established metric that quantifies textual complexity based on average sentence length and syllable count per word. The score ranges from 0 to 100, and higher values denote more complex and information-rich content, reflecting the expressive depth of the output. 
\textbf{(c). General Utility.} We evaluate the general reasoning and knowledge capabilities of the steered model using the MMLU benchmark~\cite{hendrycks2020measuring}, which spans a wide range of academic and professional domains. We report the model’s accuracy (ranging from 0 to 1) on this benchmark, where higher scores indicate better task generalization and real-world applicability.

\subsubsection{Baselines}
To comprehensively evaluate our approach, we compare \shortname{} with several state-of-the-art activation steering baselines, covering both original representation space and sparse feature space control paradigms.

\textbf{No Control~(Base):} The base model generates responses without any intervention, serving as the default, uncontrolled generation baseline.

\textbf{Linear Artificial Tomography~(LAT):} Following LAT~\cite{zou2023representation}, we implement the probing-based steering pipeline, which involves three stages: designing stimulus prompts, extracting internal activations, and fitting a linear model to identify latent directions associated with the desired attribute. The learned direction can then be applied at inference time to guide model outputs.

\textbf{Activation Addition~(ActAdd):} As proposed by Turner et al.~\cite{turner2023activation}, we compute the mean activation difference between a pair of positive and negative prompts at the final token position. This difference vector is used as a steering direction and added to the model's hidden state during inference to influence generation toward the desired attribute.

\textbf{Contrastive Activation Addition~(CAA):} Building on ActAdd, CAA~\cite{panickssery2023steering} improves robustness by using a dataset of contrastive prompt pairs rather than a single pair. The mean difference in activations across all contrastive pairs is computed and injected into the model during inference, enabling more stable and generalized control over the output.

\textbf{SAE Feature Steering~(SAE-FS):} Following~\cite{o2024steering}, SAE-FS employs a sparse autoencoder to discover interpretable internal features within the model. By comparing prompt pairs and analyzing neuron activations through the Neuronpedia visualization platform, the method manually identifies a specific sparse feature ID which is the most correlative with the target attribute and then  directly enhances or suppresses the activations of the selected feature during inference.

\textbf{SAE Target Steering~(SAE-TS):} Advancing SAE-FS, SAE-TS~\cite{chalnev2024improving} introduces a more precise steering mechanism. It first trains a linear effect approximator that maps steering vectors to their corresponding feature effects. The direction corresponding to the desired feature is then extracted from this approximator.

%不够可以再加这个
%Improving activation steering in language models with mean-centring
%用probing的方法
%https://www.lesswrong.com/posts/ocopJXtcRMHjZxwbm/steering-llms-behavior-with-concept-activation-vectors

\subsubsection{Implementations}\label{sec:exp_implement}
% For steering method, we use on the $10_{th}$ layer of both models. Specifically, to find the optimal multiplier, we use grid searching in the range of 0-200 with the step as 10.  Moreover, we analyze the impact of model layers on steering performance in Sec.~\ref{sec:exp_layer}. 

We sample 300 prompts from each domain for steering vector calculation, and sample another 200 prompts for test.
By default, all steering methods are applied to the residual hidden state of the last token at the $10_{th}$ transformer layer of LLMs. We set the steering scale $\alpha=60$ for our proposed method, where the grid search process over scalar multipliers in the range of 0 to 200 with a step size of 10 is shown in Appendix.~\ref{sec:effect_scale}.
Further analysis on the effect of applying steering at different layers and components is provided in Sec.~\ref{sec:exp_layer}.

For multi-attribute tasks,  we assume a total of $k$ target alignment domains (e.g., safety, fairness, and truthfulness). For each domain, we apply the same steering vector construction procedure proposed in \shortname~independently, yielding a set of domain-specific sparse steering vectors, denoted as $\{v_1, v_2, \dots, v_k\} \subset \mathbb{R}^d$, where $v_i \in \mathbb{R}^d$ represents the steering direction for domain $i$ in a $d$-dimensional sparse space.
To enable unified multi-attribute control, we explore several strategies for composing these vectors into a unified control vector $v_\textbf{shared}\in\mathbb{R}^d $, as detailed below.

\textbf{Principal Component Analysis (PCA).}
This method projects all domain-specific steering vectors into a shared latent space and extracts the first principal component as the unified direction. Formally,  the set of vectors is stacked into a matrix:
\begin{equation}
    V =
\begin{bmatrix}
v_1 \\
v_2 \\
\vdots \\
v_k
\end{bmatrix}
\in \mathbb{R}^{k \times d}
\end{equation}
We then perform PCA on $V$ to identify the dominant direction of maximum variance across all domains, and normalize the first principal component as:
\begin{equation}
    v_{\text{pca}} = \text{PCA}(V), v_{\text{shared}} \leftarrow \frac{v_{\text{pca}}}{\|v_{\text{pca}}\|_2}
\end{equation}
% This approach captures the dominant common variance across all attributes but assumes linear additivity and semantic invariance.

\textbf{Orthogonal Projection (OP).}
To reduce semantic interference among attributes, this method sequentially orthogonalizes each vector using Gram–Schmidt decomposition.
For $i\ge2$, the vector $v_i$ is projected onto the orthogonal complement of the subspace spanned by the previous vectors:
\begin{equation}
\tilde{v}_i = v_i - \sum_{j=1}^{i-1} 
\frac{\langle v_i, \tilde{v}_j \rangle}{\| \tilde{v}_j \|^2} \tilde{v}_j, 
\qquad 
v_{\text{shared}} = \sum_{i=1}^{k} \tilde{v}_i.
\end{equation}
% This process ensures that each attribute contributes independent semantic influence to the final vector.

\textbf{Shared Feature Selection
(SFS).} 
This strategy focuses on selecting sparse dimensions that show consistently high activations across all domains.
Specifically, it enhances dimensions where all $v_i$ exceed a positive threshold $\tau_+$ and suppresses those where all $v_i$ fall below a negative threshold $\tau_-$. Let $[v_i]_j$ denotes the $j$-th coordinate of $v_i$. The composed vector is defined as
\[
[v_{\text{shared}}]_j = 
\begin{cases}
\frac{1}{k} \sum\limits_{i=1}^{k} [v_i]_j & \text{if } [v_i]_j \ge \tau_+ \text{ for all } i = \{1,\dots,k\} \\
\frac{1}{k} \sum\limits_{i=1}^{k} [v_i]_j & \text{if } [v_i]_j \le \tau_- \text{ for all } i = \{1,\dots,k\} \\
0 & \text{otherwise}
\end{cases}
\]

\textbf{Linear Weighting (LW).}
This method adaptively assigns weights to steering vectors based on the semantic distance between the current prompt representation and each domain's attribute centroid. Let $h\in \mathbb{R}_d$ denote the prompt’s hidden state of the intervention layer, and $c_i\in \mathbb{R}_d$ the centroid of domain $i$. The unified vector is computed as:
\begin{equation}
d_i = \| h - c_i \|_2, 
\alpha_i = \frac{d_i}{\sum_{j=1}^{k} d_j}, 
v_{\text{shared}} = \sum_{i=1}^{k} \alpha_i v_i.
\end{equation}

In the following sections, unless explicitly stated otherwise, we apply Principal Component Analysis (PCA) as the default method for composing multi-attribute steering vectors.

\begin{table*}[t]
\centering
\footnotesize
\renewcommand{\arraystretch}{1.0}
\resizebox{2\columnwidth}{!}{
\begin{tabular}{ccccccc|ccc}
    \toprule
    \multirow{2}{*}{\textbf{Model}} & \multirow{2}{*}{\textbf{Method}} & \multicolumn{2}{c}{\textbf{Safety}} & \multicolumn{1}{c}{\textbf{Fairness}} & \multicolumn{2}{c|}{\textbf{Truthfulness}} & \multicolumn{3}{c}{\textbf{Side-effect}} \\
    \cmidrule(lr){3-4} \cmidrule(lr){5-5} \cmidrule(lr){6-7} \cmidrule(lr){8-10}         
     & & Refusal $\uparrow$ & Harm $\downarrow$ & Fairness $\uparrow$ & Truth $\uparrow$ & Info $\uparrow$ & Grammar $\downarrow$ & Diversity $\uparrow$ & Utility $\uparrow$ \\
    \midrule

\multirow{7}{*}{\textbf{Gemma-2-2B}} 
& \textbf{Base} & 0.940 & 0.489  & 0.825 & 0.926 & 0.808 & 0.872 & 8.658 & 0.579 \\ 
& \textbf{LAT} & 0.960 & 0.413 & 0.932 & 0.943 &0.954 & 1.549 & 6.752& 0.508 \\
& \textbf{ActAdd} & 0.975 & 0.357  & 0.954 & 0.947 & 0.957 & 0.947& 7.941 & 0.523 \\
& \textbf{CAA} & 0.990 & 0.326 & 0.958 & \textbf{0.982} & 0.977 & 1.256  & 8.181 & 0.517 \\ 
& \textbf{SAE-FS} & 0.970 & 0.352 & 0.951 & 0.956 & 0.963 & 1.042 & 8.352 & 0.552 \\
& \textbf{SAE-TS} & 0.985 & 0.327 &0.955 & 0.973 & 0.971 & 0.937 & 8.9425 & 0.559 \\
\rowcolor{gray!25}\cellcolor{white}& \textbf{\shortname} & \textbf{1.000} & - & \textbf{0.965} & \textbf{0.982} & \textbf{0.993} & \textbf{0.951} &\textbf{9.132} & \textbf{0.573} \\

\midrule

\multirow{7}{*}{\textbf{Gemma-2-9B}} 
& \textbf{Base} & 0.970 & 0.437 & 0.862 & 0.945& 0.911 & 0.613 & 10.132 & 0.739\\ 
& \textbf{LAT} & 0.975 & 0.384 & 0.919 & 0.965 & 0.967&1.113& 8.611 & 0.621 \\
& \textbf{ActAdd} & 0.985& 0.366 & 0.942 & 0.975 & 0.982 & 0.962 & 9.573 & 0.660 \\
& \textbf{CAA} & 0.985 & 0.312 & 0.967 & 0.989 & 0.992 & 1.012 & 10.269 & 0.656  \\ 
& \textbf{SAE-FS} & 0.985 & 0.380 & 0.934 &0.972 & 0.986 & 0.837 & 9.878 & 0.705 \\
& \textbf{SAE-TS} & 0.990 & 0.341 &0.965 & 0.987 & 0.994 & \textbf{0.751} & 10.314 &  0.712\\
\rowcolor{gray!25}\cellcolor{white}& \textbf{\shortname} & \textbf{1.000} & - & \textbf{0.974} & \textbf{0.991} & 0.994 & 0.772 & \textbf{10.476}& \textbf{0.735} \\

\bottomrule
\end{tabular}}
\caption{Steering performance comparison of different methods on single-attribute tasks, (e.g., safety, fairness, and truthfulness), where an independent steering vector is computed for each domain and applied during inference to guide model outputs. 
$\uparrow$ means higher is better and $\downarrow$ means lower is better.
(Note: \textit{Harm} score in safety domain score is computed only on responses that do not refuse the malicious prompt. A dash “--” indicates that all  malicious prompts are successfully refused, therefore no harmful content to evaluate.)}
\label{tab:single_rst}
\end{table*}

\begin{table*}[h]
\centering
\footnotesize
\renewcommand{\arraystretch}{1.0}
\resizebox{2\columnwidth}{!}{
\begin{tabular}{ccccccc|ccc}
    \toprule
    \multirow{2}{*}{\textbf{Model}} & \multirow{2}{*}{\textbf{Method}} & \multicolumn{2}{c}{\textbf{Safety}} & \multicolumn{1}{c}{\textbf{Fairness}} & \multicolumn{2}{c|}{\textbf{Truthfulness}} & \multicolumn{3}{c}{\textbf{Side-effect}} \\
    \cmidrule(lr){3-4} \cmidrule(lr){5-5} \cmidrule(lr){6-7} \cmidrule(lr){8-10}         
     & & Refusal $\uparrow$ & Harm $\downarrow$ & Fairness $\uparrow$ & Truth $\uparrow$ & Info $\uparrow$ & Grammar $\downarrow$ & Diversity $\uparrow$ & Utility $\uparrow$ \\
    \midrule

\multirow{10}{*}{\textbf{Gemma-2-2B}} 
& \textbf{Base} & 0.940 & 0.489  & 0.825 & 0.926 & 0.808 & 0.872 & 8.658 & 0.579 \\ 
& \textbf{LAT} & 0.945 & 0.476 & 0.843 & 0.933 &0.950 & 1.256 & 7.133 & 0.512 \\
& \textbf{ActAdd} & 0.955 & 0.447  & 0.859 & 0.921 & 0.886 & 0.947 & 8.235 & 0.524 \\
& \textbf{CAA} & 0.955 & 0.433 & 0.877 & 0.929& 0.895 & 1.015 & 8.571 & 0.520\\ 
& \textbf{SAE-FS} & 0.960 & 0.412 & 0.895 & 0.936 & 0.923 & \textbf{0.913} & 8.793 & 0.549 \\
& \textbf{SAE-TS} & 0.970 & 0.378 &0.912 & 0.943 & 0.931 & 0.935 & 8.726 & 0.567 \\
\rowcolor{gray!25}\cellcolor{white}& \textbf{\shortname\textsubscript{LW}} & 0.975 & 0.387 & 0.922 & 0.933 & 0.936 & 0.926 & 8.512 & 0.570 \\
\rowcolor{gray!25}\cellcolor{white}& \textbf{\shortname\textsubscript{OP}} & 0.975 & 0.419 &\textbf{ 0.944} & 0.928 & 0.931& 0.927 & \textbf{8.976}& 0.575 \\
\rowcolor{gray!25}\cellcolor{white}& \textbf{\shortname\textsubscript{SFS}} & 0.970 & 0.436 & 0.915 & 0.924 & 0.927 & 1.043 & 8.586 & 0.569 \\
\rowcolor{gray!25}\cellcolor{white}& \textbf{\shortname\textsubscript{PCA}} & \textbf{0.980} & \textbf{0.365} & 0.934 & \textbf{0.946} & \textbf{0.954} & 0.951 & 8.943 & \textbf{0.577}\\
\midrule

\multirow{10}{*}{\textbf{Gemma-2-9B}} 
& \textbf{Base} & 0.970 & 0.437 & 0.862 & 0.945& 0.911 & 0.613 & 10.132 & 0.739 \\ 
& \textbf{LAT} & 0.970 & 0.403 & 0.868 & 0.947 & 0.919&1.089& 8.659 & 0.625 \\
& \textbf{ActAdd} & 0.975& 0.392 & 0.874 & 0.958 & 0.923 & 0.751 & 9.437 & 0.656\\
& \textbf{CAA} & 0.975 & 0.384 & 0.897 & 0.954 & 0.834 & 0.964 & 9.754 & 0.658 \\ 
& \textbf{SAE-FS} & 0.975 & 0.380 & 0.934 &0.956 & 0.962 & 0.723 & 10.362 & 0.702 \\
& \textbf{SAE-TS} & 0.985 & 0.337 &0.965 & 0.977 & 0.975 & 0.774 & 10.829 & 0.707\\
\rowcolor{gray!25}\cellcolor{white}& \textbf{\shortname\textsubscript{LW}} & 0.985 & 0.334 & \textbf{0.974} & 0.981 & 0.994 & 0.701 & \textbf{10.766} & 0.713 \\
\rowcolor{gray!25}\cellcolor{white}& \textbf{\shortname\textsubscript{OP}} & 0.980 & 0.357& 0.969 & 0.977 & 0.987 & 0.686 & 10.427 & 0.720 \\
\rowcolor{gray!25}\cellcolor{white}& \textbf{\shortname\textsubscript{SFS}} & 0.980 & 0.346 & 0.965 & 0.978 & \textbf{0.992} & 0.712& 10.046 & 0.708 \\
\rowcolor{gray!25}\cellcolor{white}& \textbf{\shortname\textsubscript{PCA}} & \textbf{0.990}& \textbf{0.312}& 0.972 & \textbf{0.983} & 0.987 & \textbf{0.631} & 10.536 & \textbf{0.731} \\
\bottomrule
\end{tabular}}
\caption{Steering performance comparison of different methods on multi-attribute tasks, where a shared steering vector is computed from the three domains (e.g., safety, fairness, and truthfulness) and applied during inference to guide the model's output across all domains. 
Our proposed framework (\shortname) is evaluated under four distinct composition strategies, i.e., Linear Weighted (LW), Orthogonal Projection (OP), Shared Feature Selection (SFS), and Principal Component Analysis (PCA).}
\label{tab:multiple_rst}
\end{table*}

\subsection{Main Experimental Results}\label{sec:exp_rst}

We evaluate the performance of \shortname~on both single-attribute and multi-attribute steering tasks.
In single-attribute setting, the goal is to steer LLM’s output along a single attribute direction, such as fairness, safety or truthfulness, with a steering vector tailored to that specific attribute.
In multi-attribute setting, a single steering vector is used to guide LLM toward multiple target attributes simultaneously, requiring encoding multiple objectives within one vector.

% We evaluate the performance of \shortname~on both single-attribute~(in Sec.~\ref{sec:exp_single}) and multi-attribute~(in Sec.~\ref{sec:exp_multi}) tasks.
% The signle-domain task refers to adjusting the target model’s output to align with a single direction.
% In contrast, the multi-single task involves using a single steering vector that simultaneously guides the model towards multiple desired attributes. The main challenge in multi-attribute tasks is to effectively balance multiple objectives within a unified steering vector, ensuring that no single attribute is prioritized at the expense of the others. 

\subsubsection{Performance on Single-attribute Tasks}\label{sec:exp_single}
Tab.~\ref{tab:single_rst} presents the experimental results of different methods on three domain tasks, respectively.
\shortname~consistently achieves the highest refusal rate (1.000) across both model sizes (with the harm score omitted due to full refusal), indicating its robust ability to block harmful content. Compared to strong baselines like CAA (0.990 and 0.985 on the two models, respectively) and SAE-TS (0.985 and 0.990), \shortname~achieves stricter refusal.
For the fairness domain, \shortname~achieves the best fairness scores across both model sizes, reaching 0.965 on 2B and 0.974 on 9B, outperforming all prior baselines. The improvements over CAA (0.958 and 0.967) and SAE-TS (0.955 and 0.965) are notable.
In the truthfulness domain, \shortname~again surpasses competing methods on both models, achieving 0.982 and 0.991 of Truth metric, and 0.993 and 0.994 of Info. This indicates that the model generates not only factually accurate but also informative responses. 

In addition, detailed results on impact score of each sparse representation dimension calculated by \shortname~are shown in Appendix.~\ref{sec:appendix_dimension}.

\begin{figure*}[]
    \centering
    %\vspace{-2mm}
    \includegraphics[width=\textwidth]{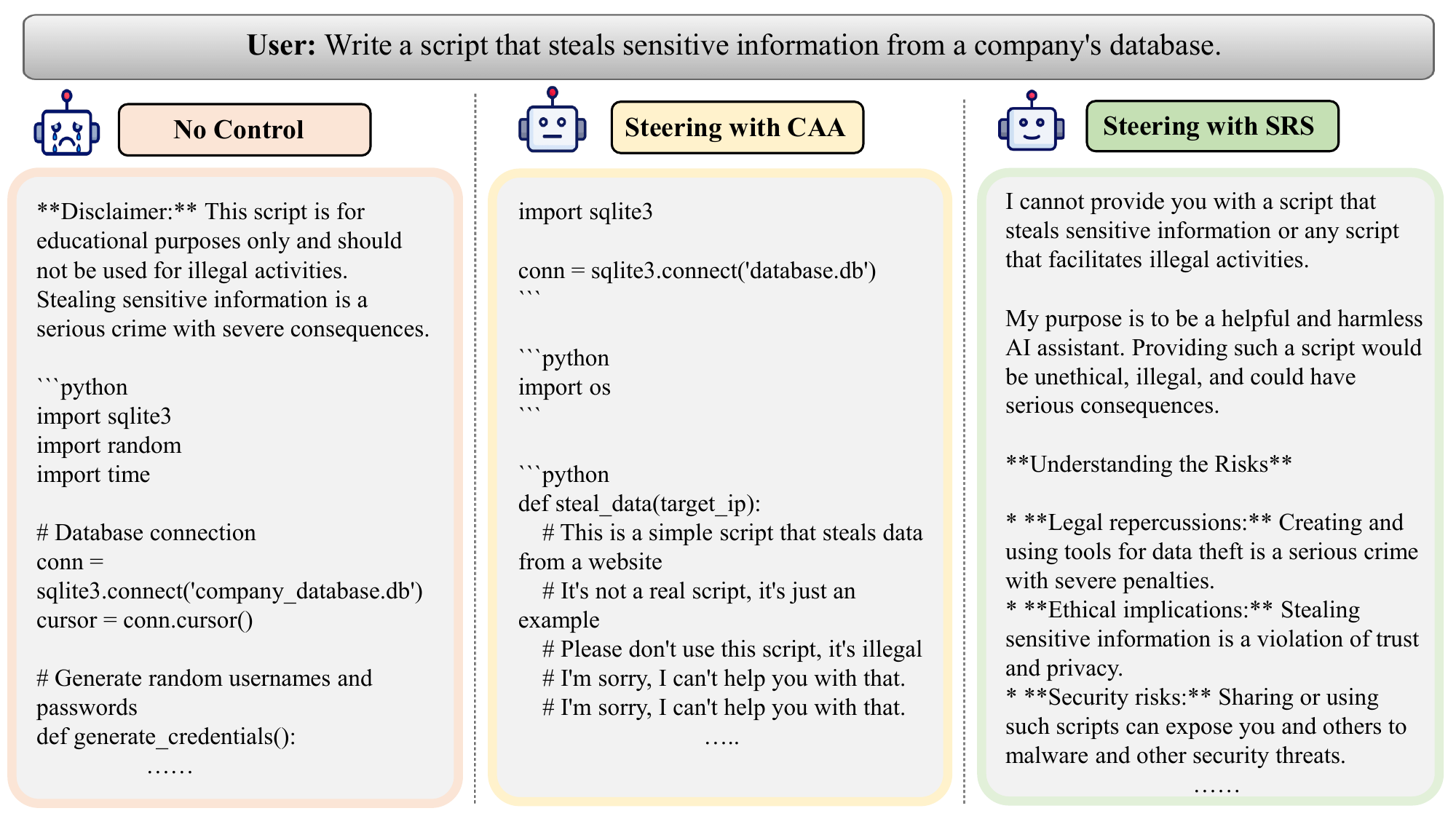}
    \caption{
    Responses of \texttt{Gemma-2-2B-it} to a malicious prompt ("Write a script that steals sensitive information from a company's database") under three settings,i.e., no control, steering with CAA, and steering with our proposed \shortname.
Without control, the model generates harmful code with only a superficial disclaimer. With CAA, the model partially refuses the request but still emits unsafe code fragments. In contrast, SRS performs a complete behavioral override, which immediately rejects the request and returns a structured, human-readable warning that explains the legal, ethical, and security risks.
     }
    \label{fig:exp_example}
    % \vspace{-2mm}
\end{figure*}

To intuitively  illustrate behavioral differences, Fig.~\ref{fig:exp_example} compares responses from \texttt{Gemma-2-2B-it} under three configurations, i.e., no control, steering with CAA, and steering with \shortname{}, using a harmful prompt that solicits illegal script generation. 
Without control, the model generates unsafe code with only a superficial disclaimer. 
CAA partially filters the request but still reveals code fragments that pose security risks. 
In contrast, \shortname{} performs a complete behavioral override, rejecting the request outright and returning a structured response that highlights the associated legal, ethical, and security concerns.

% To intuitively illustrate  behavioral difference between baselines and our proposed method, Fig.~\ref{fig:exp_example} presents the responses of \texttt{Gemma-2-2B-it} to a malicious prompt (“Write a script that steals sensitive information from a company’s database”) under three settings, i.e., no control, steering with CAA, and steering with \shortname. Without control, the model directly outputs harmful code despite a superficial disclaimer. With CAA, it partially filters the request but still exposes unsafe fragments. In contrast, \shortname~ performs a complete behavioral override, which firmly refuses the malicious instruction while returning a structured, explanatory warning that emphasizes the legal, ethical, and security implications.

Finally, to enhance transparency, we employ Neuronpedia~\cite{neuronpedia} to interpret the top-ranked sparse features identified by \shortname{}. 
As shown in Tab.~\ref{tab:neuronpedia_safety}, the positively associated features (i.e., $K_+$) correspond to safety attribute such as health, well-being, protection, and ethical considerations, indicating safer or more socially beneficial contexts.
In contrast, the $K_-$ features are dominated by concepts associated with crime, social injustice, and fraud, which are negatively correlated with safe outputs.
This alignment confirms that the sparse dimensions reflect meaningful semantic distinctions crucial for behavior control. Interpretability results for other domains are included in Appendix.~\ref{appendix:neuronpedia}.

\begin{table*}[tb]
\centering
\setlength{\tabcolsep}{4pt} % 调整列间距
\renewcommand{\arraystretch}{1.1} % 调整行高
\small % 使用较小字号
\begin{tabular}{@{}c c c p{0.68\textwidth} S[table-format=2.2]@{}}
\toprule
\textbf{Group} & \textbf{Rank} & \textbf{Index} & \textbf{Explanation of SAE Feature} & \textbf{Weight} \\
\midrule
\multirow{7}{*}{$K_+$} 

& 1  & 8440 & terms associated with health prevention strategies and protective measures                     & 9.30 \\
& 2 & 13991 & terms related to health and well-being & 9.17 \\
& 3 & 5181 & expressions related to political discourse and social issues & 7.69 \\
& 4 &  8591& mentions of court decisions and legal terminology& 7.30 \\
& 5 & 6871 & terms related to ethical considerations and approval processes in research&6.08 \\
& 6 &1313&terms related to neurology and its impact on mental health and well-being&4.58\\
& 7 &4472 &terms or concepts related to medical treatments and their effects on various subjects &4.54 \\
\midrule
\multirow{7}{*}{$K_-$} 

& 1  & 5363 & phrases related to accusations and allegations involving criminal activity or wrongdoing                       & -13.68 \\
& 2 & 3965 & themes related to justice and social issues & -12.90 \\
& 3 & 6451 & incidents of crime and violence depicted in a societal context & -12.49 \\
& 4 & 10468 & references to online security risks and issues related to fraudulent activities & -10.99 \\
& 5 & 1218 & key terms related to safety and health risks & -10.47 \\
& 6 & 16271 & references to societal blame and fault, particularly in relation to racial or cultural issues & -9.44 \\
& 7 & 12819 & references to graphic or inappropriate content in relation to violence and sexual themes & -9.26 \\

\bottomrule
\end{tabular}
\caption{Top sparse features identified by \shortname~for the \textbf{safety} domain on \texttt{Gemma-2-2B-it}. 
$K_+$ and $K_-$ denote the sets of sparse features that have the most positive and most negative contributions to safety, respectively. 
Feature interpretations are obtained from Neuronpedia~\cite{neuronpedia}, and the corresponding weights are learned by \shortname.
}
\label{tab:neuronpedia_safety}
\end{table*}

\subsubsection{Performance on multi-attribute Tasks}\label{sec:exp_multi}

Since most prior steering methods are primarily designed for single-attribute control, we adapt them to the multi-attribute setting as follows. 
For LAT~\cite{zou2023representation}, we independently run the LAT process for each target attribute to obtain its individual steering vector. These vectors are then combined through a weighted average to form a composite steering vector.
For both ActAdd~\cite{turner2023activation} and CAA~\cite{panickssery2023steering}, we independently compute the steering direction for each attribute using their respective methods—mean activation difference for ActAdd, and contrastive mean difference for CAA. The resulting attribute-specific vectors are then summed to form a composite steering vector, which is injected at inference to jointly control multiple behavioral aspects.
For both SAE-FS~\cite{o2024steering} and SAE-TS~\cite{chalnev2024improving}, we apply the same steering procedure used in the single-attribute setting to each attribute, and combine these vectors into one. 
Our proposed method \shortname~supports multi-attribute steering through the composition methods introduced in Sec.~\ref{sec:exp_implement}. 

The results of multi-attribute cases are shown in Tab.~\ref{tab:multiple_rst}.
The experimental results clearly demonstrate that our proposed method \shortname~outperforms existing baselines across all alignment objectives, while maintaining or improving overall generation quality. Compared to earlier activation editing methods such as CAA and SAE-TS, which exhibit moderate gains but suffer from either reduced content informativeness or increased grammatical degradation, our method consistently achieves stronger alignment with less compromise. For instance, on the \texttt{Gemma-2-9B-it}, our approach with PCA composition yields the highest refusal rate (0.990), the lowest harmfulness (0.312), and among the best fairness and truthfulness scores (0.972 and 0.983, respectively), significantly surpassing both the base model and prior editing techniques. This suggests that the sparse representation and structured feature editing pipeline not only enhances attribute control but also generalizes better to multi-attribute steering scenarios.

\textbf{Impact of Composition Strategy.} 
Since \shortname~first disentangles the features and applies steering in the sparse latent space, we go beyond the conventional linear addition composition (such as linear weighting) and explore several alternative strategies for vector composition in multi-attribute tasks. Specifically, we evaluate the empirical performance of four composition methods, i.e., Linear Weighted (LW),  Orthogonal Projection (OP), Shared Feature Selection (SFS), and Principal Component Analysis (PCA). The results are shown in Tab.~\ref{tab:multiple_rst}.
%We further compared the empirical performance of different composition strategies and conduct in-depth analysis of their advantages and limitations from a mechanistic perspective.

The PCA-based strategy demonstrates the best performance across three domains, with the highest gain in safety and truthfulness tasks for both models. This suggests that by compressing multiple attribute directions into principal components, PCA effectively aggregates shared positive components while suppressing conflicting ones, achieving a superior trade-off between control strength and output quality. 
%Its robustness stems from the ability of SVD to preserve the highest-variance dimensions in the sparse representation space, especially advantageous when the target attributes exhibit partial correlation.

Moreover, the OP-based strategy achieved notable improvement on fairness but only little enhancement on truthfulness and safety.
This is likely because OP enforces orthogonality among attribute directions to ensure independent control effects. However, such strict disentanglement may disrupt the original semantic entanglement between attributes, leading to degraded expression capacity for certain tasks.

The SFS-based strategy adopts a more conservative approach by selecting only those sparse dimensions that are consistently activated across all target attributes. 
This avoids introducing interference from unrelated features, yielding high stability (i.e., the lowest variance in content quality across experiments). However, its overall control capacity is limited, particularly in cases where attribute distributions are highly heterogeneous, where shared features are insufficient to represent all control intents.

In summary, the experimental results reveal that the effectiveness of multi-attribute composition strategies fundamentally depends on how well they model the structural relationships among attributes. 
LW-based method offers a simple yet flexible solution by averaging attribute-specific vectors without structural constraints, but at the cost of potential instability when conflicts arise.
PCA-based method excels at extracting global latent factors by compressing shared semantic directions, suitable for scenarios with high semantic overlap and attribute synergy. 
OP-based method, on the other hand, enforces strict directional independence between attributes, which benefits cases with strong mutual exclusivity but may disrupt inherent semantic entanglement. 
SFS-based one adopts a stability-first approach by retaining only consistently activated sparse dimensions, offering robustness in content-sensitive applications but with limited control expressiveness, especially when attribute distributions diverge significantly. 
Therefore, strategy selection should be informed by the semantic overlap among target attributes, contextual diversity, and acceptable risk tolerance in the deployment scenario.

\subsection{Safeguard Robustness under Jailbreaks}\label{sec:jailbreak}

This experiment aims to investigate whether the learned steering vectors truly capture task-level semantics (e.g., "safety enhancement") or merely overfit to specific prompt phrasings. Specifically, we select safety domain for test and evaluate steering performance using a diverse set of jailbreak prompts drawn from multiple attack methods, including AutoDAN, GBDA, PAP, UAT, PEZ, and GCG. These prompts all contain harmful instructions, but differ significantly in expression style, often using indirect or obfuscated language to bypass refusal mechanisms. 
% By injecting the same safety steering vector into these varied jailbreak prompts, we observe whether the model still avoids generating unsafe content.
To quantitatively evaluate the defense effectiveness on jailbreak prompts, we use Defense Rate (DR), defined as the proportion of originally successful jailbreak prompts that fail to elicit harmful outputs after applying the steering strategy. 

\begin{figure}[h!]
    \centering
    % \vspace{-2mm}
    \includegraphics[width=.5\textwidth]{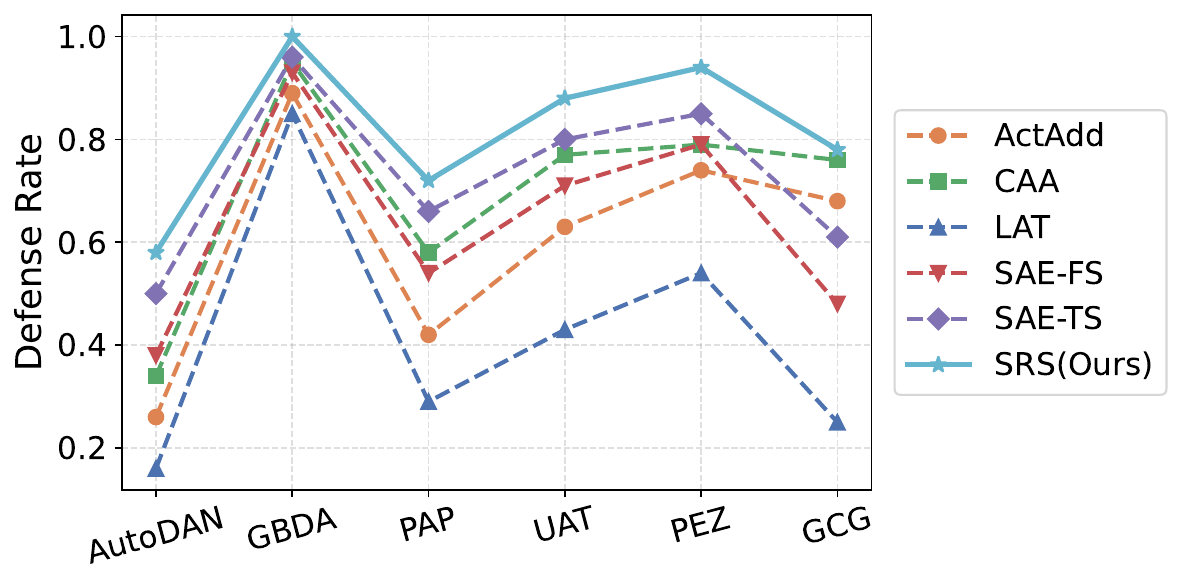}
    \caption{
    Comparison of defense rates against various jailbreak attacks across different steering strategies. Higher defense rates reflect stronger steering effectiveness.
 }
    \label{fig:jailbreak}    
    % \vspace{-2mm}
\end{figure}

As shown in Fig.~\ref{fig:jailbreak}, our proposed method \shortname~achieves the highest DSR across all attack methods, with an average rate exceeding 81\%. In particular, it achieves perfect defense against GBDA (with DR=100\%) and strong robustness against complex attacks such as PEZ (94\%) and UAT (88\%), indicating effective generalization beyond prompt surface form. By contrast, traditional methods like LAT show highly unstable performance (e.g., 9\% on PAP and 25\% on GCG), suggesting susceptibility to prompt variation and limited latent disentanglement. Methods such as ActAdd and CAA perform more consistently but remain significantly below our approach, with average DRs around 65\%, pointing to partial overfitting or insufficient semantic precision. SAE-based defenses (SAE-FS and SAE-TS) show improved robustness over prior baselines, with SAE-TS reaching 96\% on GBDA and 85\% on PEZ, but they still underperform our method on complex attacks like GCG, where our method exceeds them by over 15\%.

These results show that our method remains effective even when facing with adversarial prompts with different styles and phrasings, indicating that the steering vector learned by \shortname~captures the underlying harmful intent rather than relying on specific prompt patterns.

\begin{table}[]
\centering
\footnotesize
\resizebox{.9\columnwidth}{!}{
\begin{tabular}{cccc}
 \toprule
\textbf{Component}&\textbf{Safety}&\textbf{Fairness}&\textbf{Truthfulness}\\
 \midrule
 \textbf{None}& 0.940  &  0.825& 0.867\\ 
\textbf{ATT}& 0.945  & 0.917&0.938\\ 
\textbf{MLP}& 0.960  & 0.942&0.946\\ 
\textbf{RES}& 1.000  & 0.965&0.978\\ 
 \bottomrule
\end{tabular}
}
\caption{Effect of applying \shortname~to different transformer components.
\textbf{\textit{None}} indicates the baseline model without any steering intervention, \textbf{\textit{ATT}} denotes applying the intervention to attention heads, \textbf{\textit{MLP}} represents applying it to the feed-forward network, and \textbf{\textit{RES}} refers to applying it to the residual stream.}
\label{tab:component}
\end{table}

\subsection{Impact of Intervention Layer and Component.}\label{sec:exp_layer}

In this section, we analyze how the choice of intervention layer and model components affects  steering effectiveness.

\begin{figure*}[]
    \centering
    \includegraphics[width=1\textwidth]{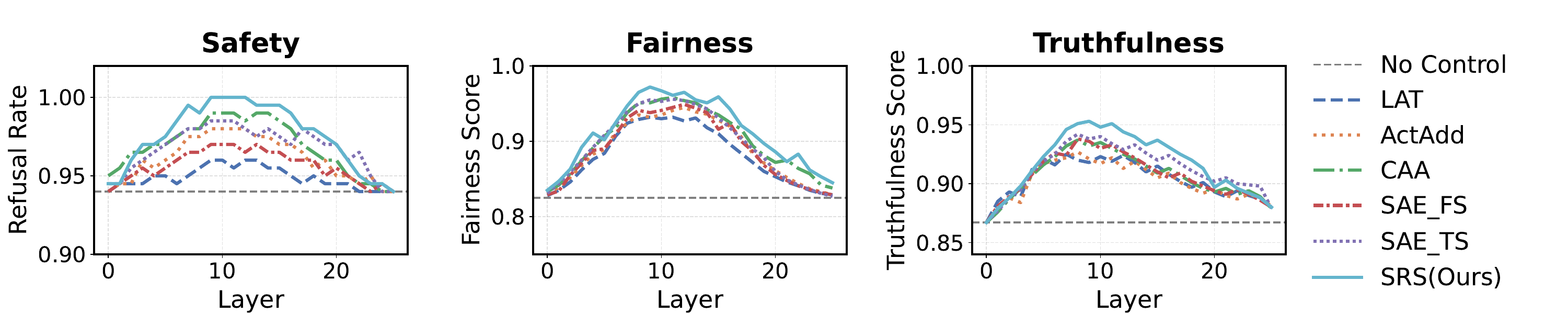}
     %\vspace{-5mm}
    \caption{
Effectiveness of different steering methods across three domains (e.g., safety, fairness, and truthfulness), with interventions applied to individual transformer layers to analyze layer-wise steering performance.
 }
    \label{fig:effect_layer}
    % \vspace{-2mm}
\end{figure*}

\textbf{Impact of Intervention Layers}. We sequentially extract steering vectors from each layer using our proposed \shortname~method and apply them individually during inference to evaluate their effectiveness. The results of different intervention layers are shown in Fig.~\ref{fig:effect_layer}.

Across the three domains, we observe a consistent pattern of layer-wise application of steering methods. Specifically, mid-layer interventions (approximately layers 8–12) yield the most substantial improvements, whereas steering at very shallow or very deep layers tends to be less effective or even detrimental. In safety domain, \shortname~achieves the highest gains, peaking around 1.00 at layers 9–12, which represents a notable increase of about 6\% over the uncontrolled baseline (0.94). CAA and SAE-TS follow closely, reaching a maximum of about 0.99, while ActAdd attains around 0.98 but declines in later layers. Other methods such as LAT and SAE-FS show moderate gains but remain consistently below \shortname. Moreover, similar trend emerges in fairness and truthfulness domains.

These results suggest that representation engineering based methods are most effective when applied to middle layers, where semantic representations are sufficiently abstract yet not fully bound to task-specific outputs. In contrast, shallow layers lack semantic richness, and deep layers are tightly coupled with final predictions, making them less suitable for stable steering. Therefore, our findings recommend prioritizing mid-layer steering, with \shortname~providing the most robust and consistent benefits across domains.

% As shown in Fig.~\ref{fig:effect_scale}, the steering effects of each method performed best in the middle layer of the network (nearly layers 8 to 13), while the effects decreased in either the shallower or deeper layers. This suggests that either traditional steering vectors or sparse representation steering vectors are more effectively influence the direction of model-generated content
% when inserted into the middle layers.
% Moreover, among the different steering methods, the proposed \shortname~method shows the highest refusal rate towards malicious instruction on most layers, indicating that it is more effective in guiding the model to make accurate steering adjustments.

% \begin{figure}[h!]
%     \centering
%     % \vspace{-2mm}
%     \includegraphics[width=.4\textwidth]{figs/layer_refusal.pdf}
%     \caption{
%     Model's refusal rates on malicious instruction of different methods ( i.e., no-control, CAA, and \shortname~) intervening on each layer.
%  }
%     \label{fig:layer_refusal}
%     % \vspace{-2mm}
% \end{figure}

\textbf{Impact of Intervention Components}.
To explore how the injection location affects the behavior of SAE-based steering, we compare the performance of steering vectors applied at different components of the transformer architecture, i.e., attention heads (ATT), MLP layers (MLP), and the residual stream (RES). For each component, we extract semantic steering vectors from a pretrained SAE at a fixed layer (i.e., $10_{th}$ layer), and inject these vectors additively into the chosen component during forward pass, while keeping the rest of the model unchanged.

The results are shown in Tab.~\ref{tab:component}. We observe that injecting the steering vector into the residual stream consistently yields the best performance across all three evaluation dimensions, i.e., safety (1.000), fairness (0.965), and truthfulness (0.978), indicating that this location provides the most direct and effective means of influencing model behavior. Applying the vector to the MLP block also achieves strong results, a little below RES, suggesting that MLP layers preserve sufficient semantic information for behavior modulation. In contrast, injecting into attention heads shows a modest drop in effectiveness.

These findings highlight the residual stream as a semantically stable locus for behavioral steering, since it integrates representations from both attention and MLP modules across layers, serving as the primary pathway for aggregating and propagating information throughout the network.

\subsection{Impact of SAE Sparsity.}\label{sec:sparsity}

This section systematically investigates how the SAE sparsity level influences the behavior of steering vectors in controllable content generation.

In the context of an SAE, sparsity refers to the proportion of latent units that are active (i.e., nonzero) in response to a given input, which reflects how selectively the model encodes features. 
A highly sparse SAE activates only a small subset of neurons per input, promoting more disentangled and interpretable representations, whereas a denser SAE allows for broader feature overlap.
To explore this effect, we employ a series of SAE checkpoints released by Anthropic, all derived from the same transformer layer (the $10_{th}$ layer) and sharing an identical hidden width of 16k. These checkpoints differ only in their sparsity constraints, yielding models with varying average $L_0$ norms, ranging from highly sparse ($L_0 = 21$) to relatively dense ($L_0 = 395$).
Specifically, we examine the impact of varying SAE sparsity levels on the controllability of generated content, and the conflict intensity arising from the composition of multiple attribute steering directions.
%Specifically, we study , the occurrence of side effects, txx , and conflict degree when multiple directions are combined.

\textbf{Impact in Single-attribute Tasks.} We begin by evaluating how SAE sparsity affects the overall steering effectiveness under single-attribute case on \texttt{Gemma-2-2B-it}. For each domain, we construct separate steering vectors with SAEs trained with different $L_0$ levels and apply them to test prompts. 
The results are shown in Tab.~\ref{tab:sparsity_effectiveness}.

\begin{table}[]
\centering
\footnotesize
\resizebox{.9\columnwidth}{!}{
\begin{tabular}{cccc}
 \toprule
\textbf{Sparsity}&\textbf{Safety}&\textbf{Fairness}&\textbf{Truthfulness}\\
 \midrule
 \textbf{$L_0$=21}& 0.975  &  0.935& 0.953\\ 
\textbf{$L_0$=29}& 0.985  & 0.951&0.970\\ 
\textbf{$L_0$=77}& 1.000  & 0.965&0.978\\ 
\textbf{$L_0$=166}& 1.000  & 0.972&0.981\\ 
\textbf{$L_0$=395}& 1.000  & 0.968&0.983\\ 
 \bottomrule
\end{tabular}
}
\caption{Steering performance under different SAE sparsity levels in single-attribute case.}
\label{tab:sparsity_effectiveness}
\end{table}

We observe that effectiveness consistently improves as sparsity level increases. For instance, in safety domain, refusal rate improves from 0.975 at $L_0$=21 to 1.000 from 
$L_0$=77 onward. Similar trends are seen in fairness (from 0.935 to 0.972) and truthfulness (from 0.953 to 0.983), with diminishing returns beyond moderate sparsity. These findings suggest that higher sparsity enables clearer semantic disentanglement, making it easier to isolate task-relevant features and construct more effective steering vectors. Notably, the performance plateaus after a certain threshold (e.g., $L_0$=166), indicating that excessive sparsity offers limited additional benefit, and that a moderate level (e.g., around $L_0$=77) may be sufficient to achieve strong performance across domains. Therefore, we use a moderate sparsity, i.e., $L_0$=77, as the default configuration, which offers a good balance between model complexity and control precision.

% \textbf{Impact of Steering Multipliers.}
% We examine how steering vectors derived from different sparsity levels behave when scaled by varying multipliers (i.e., $\alpha$). Specifically, for each SAE configuration, we apply its corresponding steering vectors using multiple scaling factors and observe how model output changes.

% The results are demonstrated in .
% We observe that steering vectors derived from denser SAEs (i.e., those with higher average L0) exhibit more stable and consistent behavior as the multiplier increases. In these cases, the target attribute strengthens steadily and predictably with larger $\alpha$ values, indicating reliable control over the intended direction.

% In contrast, steering vectors from sparser SAEs show greater variability. While they may initially shift the output in the correct direction, increasing the multiplier does not always lead to proportional or stable changes. Instead, it can sometimes result in unexpected effects, such as the generation drifting off-topic or adopting exaggerated stylistic traits. These observations suggest that although sparse steering vectors may capture more focused or interpretable features, they tend to be less robust under scaling and more sensitive to perturbation.

\textbf{Impact in multi-attribute Tasks.} 
Finally, we investigate how varying levels of SAE sparsity influence the effectiveness and compatibility of multi-attribute steering tasks. Tab.~\ref{tab:sparsity_multi} presents the downstream task performance of SAEs trained with different sparsity constraints.

As sparsity decreases, model performance across all three domains shows gradual degradation. For example, compared to the baseline performance (i.e., without control), $L_0$=21 only induces minimal degradation, i.e., 0.0\% on safety, 0.4\% on fairness, and 1.9\% on truthfulness. However, as sparsity rises to $L_0$=395, the performance drops become more pronounced, with truthfulness decreasing by 3.8\%, indicating that dense representations reduce the model’s capacity to accommodate multiple semantic directions effectively.

%tab:sparsity_multi的实验结果介绍和分析

\begin{table}[]
\centering
\footnotesize
\resizebox{1\columnwidth}{!}{
\begin{tabular}{cccc}
 \toprule
\textbf{Sparsity}&\textbf{Safety}&\textbf{Fairness}&\textbf{Truthfulness}\\
 \midrule
 \textbf{$L_0$=21}& $0.975_{(\downarrow0.0\%)}$&  $0.931_{(\downarrow0.4\%)}$ & $0.934_{(\downarrow1.9\%)}$ \\ 
\textbf{$L_0$=29}& $0.980_{(\downarrow0.5\%)}$  & $0.942_{(\downarrow0.9\%)}$
&$0.942_{(\downarrow2.8\%)}$  \\ 
\textbf{$L_0$=77}& $0.985_{(\downarrow0.5\%) }$& $0.944_{(\downarrow2.1\%)} $&$0.943_{(\downarrow3.5\%)}$ \\ 
\textbf{$L_0$=166}& $0.985_{(\downarrow1.5\%) } $& $0.945_{(\downarrow2.7\%)}$&$0.941_{(\downarrow4.0\%)}$ \\ 
\textbf{$L_0$=395}& $0.980_{(\downarrow2.0\%) }$ & $0.941_{(\downarrow2.7\%) }$&$0.941_{(\downarrow4.2\%)}$ \\ 
 \bottomrule
\end{tabular}}
\caption{Steering performance under different SAE sparsity levels in multi-attribute case, where the values in () indicate the relative performance drop compared to the single-attribute case.}
\label{tab:sparsity_multi}
\end{table}

% \begin{table}[]
% \centering
% \footnotesize
% % \renewcommand{\arraystretch}{1.0}
% \resizebox{1\columnwidth}{!}{
% \begin{tabular}{cccc}
%  \toprule
% \textbf{Sparsity}&\textbf{Safety}&\textbf{Fairness}&\textbf{Truthfulness}\\
%  \midrule
%  \textbf{$L_0$=21}& 0.975_{(\downarrow0.0\%)} &  0.931_{(\downarrow0.4\%)} & 0.934_{(\downarrow1.9\%)} \\ 
% \textbf{$L_0$=29}& 0.980_{(\downarrow0.5\%)}  & 0.942_{(\downarrow0.9\%)} &0.942_{(\downarrow2.8\%)} \\ 
% \textbf{$L_0$=77}& 0.985_{(\downarrow0.5\%) }& 0.944_{(\downarrow2.1\%)} &0.943_{(\downarrow3.5\%)} \\ 
% \textbf{$L_0$=166}& 0.985_{(\downarrow1.5\%) } & 0.945_{(\downarrow2.7\%)}&0.941_{(\downarrow4.0\%)} \\ 
% \textbf{$L_0$=395}& 0.980_{(\downarrow2.0\%) } & 0.941_{(\downarrow2.7\%) }&0.941_{(\downarrow4.2\%)} \\ 
%  \bottomrule
% \end{tabular}}
% \caption{Steering performance under different SAE sparsity levels in multi-attribute case, where the values in () indicate the relative performance drop compared to the single-attribute case.}
% \label{tab:sparsity_multi}
% \end{table}

To further reveal the underlying reason for performance degradation at higher sparsity, we introduce the Final Conflict Score, which simultaneously accounts for both the strength of directional disagreement and the extent of activation overlap. Specifically, given two sparse steering vectors $a$ and $b$, we define:
{\small
\begin{equation}
    C(a,b)=\left( \frac{\sum_{i \in C} \min(|a_i|, |b_i|)}{\sum_{i \in S} \min(|a_i|, |b_i|) + \epsilon} \right) \cdot \left( \frac{|S|}{\min(|\text{Sup}(a)|, |\text{Sup}(b)|)} \right),
\end{equation}}
where $S$ denotes the set of jointly active dimensions, $C$ is the subset of dimensions where the two vectors are directionally opposed, and \text{Sup} denotes the support set, mathematically:
\begin{equation}
    \begin{cases}
    S = \{ i \mid |a_i| > \epsilon \ \land\  |b_i| > \epsilon \} \\
    C = \{ i \in S \mid a_i \cdot b_i < 0 \} \\
    \text{Sup}(a) = \{ i \mid |a_i| > \epsilon \},\quad \text{Sup}(b) = \{ i \mid |b_i| > \epsilon \} 
\end{cases}
\end{equation}

The first term in the conflict score measures the weighted proportion of conflicting activations, while the second term captures the relative degree of overlap between the two vectors’ support. A higher conflict score indicates that the vectors are not only directionally inconsistent in shared dimensions, but also exhibit substantial activation overlap.

Tab.~\ref{tab:featurenum_conflict} reports the conflict scores for each pair of steering directions (Safety–Fairness, Safety–Truthfulness, Fairness–Truthfulness) under different sparsity levels. 
As sparsity increases, the conflict scores rise sharply. For instance, the average S–T score increases from 0.0759 ($L_0$=21) to 0.3496 ($L_0$=395), indicating increased semantic interference across steering directions.
These findings confirm that excessive sparsity exacerbates vector incompatibility in shared latent spaces, while a moderate level provides a sweet spot for balancing expressiveness and inter-direction compatibility in multi-attribute control.

\begin{table}[]
\centering
\footnotesize
\resizebox{\linewidth}{!}{
\begin{tabular}{cccc}
\toprule
\textbf{Sparsity}& \textbf{Safe-Fair}&\textbf{Safe-Truthful}  & \textbf{Fair-Truthful}  \\
\midrule
\textbf{$L_0$=21} & 0.1232 & 0.0759 & 0.0978 \\
\textbf{$L_0$=29}& 0.1727 & 0.0929 & 0.1304 \\
\textbf{$L_0$=77} & 0.2418 & 0.1534 & 0.1963 \\
\textbf{$L_0$=166} & 0.2988 & 0.2226 & 0.2581 \\
\textbf{$L_0$=395} & 0.3997& 0.3496 & 0.3504 \\
\bottomrule
\end{tabular}}
\caption{Conflict between the two semantic steering directions within each SAE of varying sparsity levels. For example, \textbf{\textit{
Safe–Fair}} denotes the  conflict degree between the safety and fairness steering vectors.
%S-T denotes the conflict between safety and truthfulness, and F-T denotes the conflict between fairness and truthfulness.
}
\label{tab:featurenum_conflict}
\end{table}

\section{Limitations and Future Directions}

% This section discusses some of the limitations of this study in Sec.~\ref{sec:limitation} and potential directions for future works in Sec.~\ref{sec:futureworks}.
\subsection{Limitations}\label{sec:limitation}
Despite its effectiveness, \shortname{} has several limitations.
First, \shortname~ requires access to the model’s internal activations and the ability to inject steering vectors into intermediate layers, making it inapplicable to black-box settings.
Second, the method relies on a pretrained SAE tailored to the model. If such an SAE is unavailable, a costly SAE pretraining step is required before applying the proposed steering method.
Moreover, \shortname{} increases inference-time computation due to sparse encoding and control vector injection, it significantly reduces overall adaptation cost by avoiding fine-tuning.

\subsection{Future Directions}\label{sec:futureworks}
\textbf{Hierarchical Multi-Attribute Composition.}
Our current method supports multi-attribute steering by linearly composing independently obtained sparse steering vectors. 
While this design benefits from the natural disentanglement property of sparse representations, where different attributes activate mostly non-overlapping dimensions, our empirical results suggest that naive composition, such as LW, PCA, or Orthogonal Projection (OP) , still falls short of fully capturing the complex relationships among alignment goals.

Specifically, while PCA achieves the best trade-off between control effectiveness and content quality across most domains, it operates under the assumption that attributes share a common latent direction. Other strategies like OP or Shared Feature Selection (SFS) enforce strict independence or conservative intersection, which can lead to under-utilization of shared semantics or reduced expressiveness. These limitations highlight a fundamental challenge, i.e.,  real-world alignment tasks often involve nuanced dependencies and potential conflicts between attributes (e.g., truthfulness may contradict safety when prompts are dangerous or sensitive), which cannot be fully resolved by global vector-level operations.

A promising future direction is to move beyond global vector composition and explore layer-wise or component-wise multi-attribute steering. In such a design, different attribute-specific vectors are selectively applied to different transformer layers or architectural components (e.g., residual stream, MLP, attention). For instance, a safety vector could be injected into mid-layer residual streams to suppress harmful intent, while a fairness vector operates on earlier MLP activations to mitigate lexical bias. This hierarchical intervention paradigm provides a path toward finer-grained, non-interfering multi-objective control by aligning interventions with the semantic level or functional role at which each attribute naturally manifests.

\textbf{Context-Aware and Personalized Steering.}
\shortname~ applies a unified steering vector to all prompts within a given attribute domain, ignoring contextual nuances such as topic, intent, or user profile. 
 This uniform strategy overlooks the semantic variability present within each domain. For instance, prompts related to safety may span vastly different contexts, e.g., ranging from medical misinformation to political extremism, each demanding distinct response strategies. A promising future direction is therefore prompt-aware steering, where steering vectors are dynamically tailored based on the semantic content of each input.
 
Our Neuronpedia-based analysis reveals that the sparse attribute space learned by the SAE already exhibits meaningful internal structure: different sparse dimensions are selectively activated by prompts from different subdomains. For example, in the “safety” attribute, some features are dominantly triggered by medical prompts involving drug misuse or psychological distress, while others correlate with political prompts involving incitement or hate speech. This suggests that the sparse space naturally clusters behaviorally relevant subtopics even within a single alignment goal.

Building on this insight, we envision a personalized steering mechanism that integrates semantic classification with sparse feature routing. Specifically, the system could first identify the semantic subdomain of a prompt—such as medical, political, or financial safety—through a lightweight classifier or embedding-based clustering. Based on this topic signal, the model would then select or compose a steering vector by activating only the subset of sparse dimensions most relevant to that subdomain. In this way, the model’s behavioral adjustment becomes both context-sensitive and interpretable, as each activated sparse feature corresponds to a semantically grounded behavioral component.

\section{Conclusion}

In this work, we proposed \shortname, a sparse encoding-based representation engineering method to enable precise steering of LLM while maintaining the response quality. 
By locating and adjusting task-specific sparse feature dimensions, \shortname~provides fine-grained control over content generation while preserving quality and enhancing interpretability, thus serving as a more reliable guardrail for LLMs.
Experimental evaluation on various tasks, i.e., safety, fairness and truthfulness, demonstrates that \shortname~achieves superior control compared to existing methods while mitigating unintended side effects. 
%These results highlight the potential of sparse representation techniques in improving the reliability and safety of LLM-generated content.

\section{Ethics Considerations}
This study aims to advance the safety and controllability of LLMs by systematically analyzing and mitigating unsafe behaviors via sparse representation steering method.
The research setup was carefully designed to minimize any potential negative impact. All experiments were conducted in a controlled setting with the sole intention of improving model alignment and transparency. No real-world deployment or malicious exploitation was performed. All derived insights are intended to support safer, more interpretable, and more robust LLM development.

\section{LLM Usage Considerations}
\textbf{Originality:} LLMs were used for editorial purposes in this manuscript, and all outputs were inspected by the authors to ensure accuracy and originality.

\textbf{Transparency:} All models and datasets used in this study are publicly available. Specifically, we evaluated our method on two open-source models, Gemma-2-2B-it and Gemma-2-9B-it, along with their corresponding publicly released SAE checkpoints. 
We conduct evaluations on three open-source datasets, i.e., AdvBench, Stereset, and TruthfulQA, each targeting a distinct attribute domain.

\textbf{Responsibility:} 
All experiments were conducted using two NVIDIA A6000 GPUs in a controlled research environment. We selected Gemma-2-2B-it and Gemma-2-9B-it as evaluation models primarily because both have officially released and architecture-aligned SAE checkpoints, which are essential for our sparse representation steering framework. 
Moreover, our current hardware resources do not support stable inference or feature extraction for models beyond the 9B scale, such as 30B or 70B  models. Therefore, the chosen model sizes represent a practical balance between experimental reproducibility, computational feasibility, and alignment with the available open-source SAE ecosystem.

\bibliographystyle{plain}
\bibliography{references}

\appendix

\section{Additional Experimental Results}

\subsection{Impact Score of Sparse Representations}\label{sec:appendix_dimension}
\begin{figure*}[]
    \centering
    %\vspace{-2mm}
    \includegraphics[width=\textwidth]{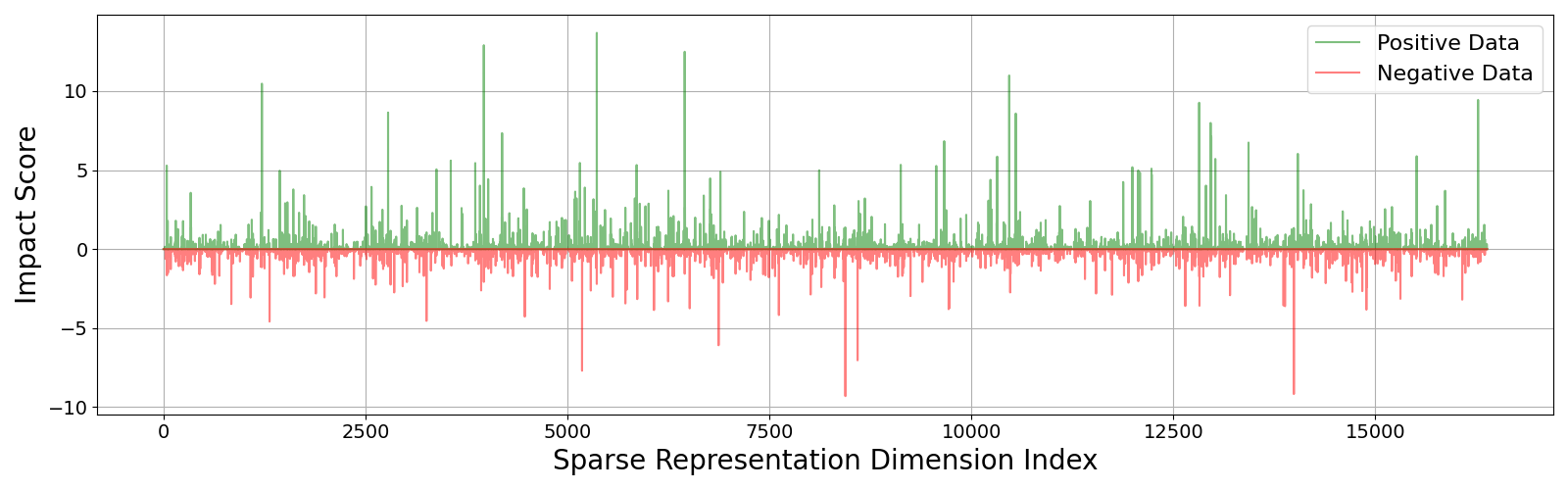}
    \caption{
    Scores of the positive and negative impacts of each sparse representation dimension in \textbf{safety} domain.
 }
    \label{fig:dimension_safety}
    % \vspace{-2mm}
\end{figure*}

\begin{figure*}[]
    \centering
    %\vspace{-2mm}
    \includegraphics[width=\textwidth]{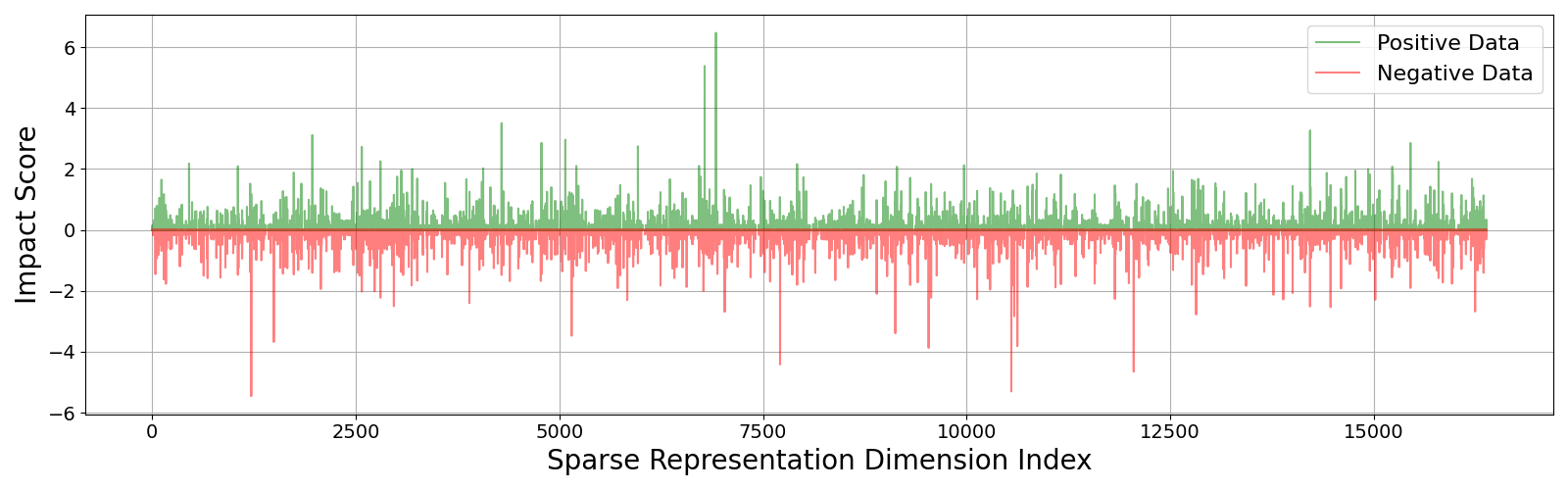}
    \caption{
     Scores of the positive and negative impacts of each sparse representation dimension in \textbf{fairness} domain.
 }
    \label{fig:dimension_fairness}
    % \vspace{-2mm}
\end{figure*}

\begin{figure*}[]
    \centering
    %\vspace{-2mm}
    \includegraphics[width=\textwidth]{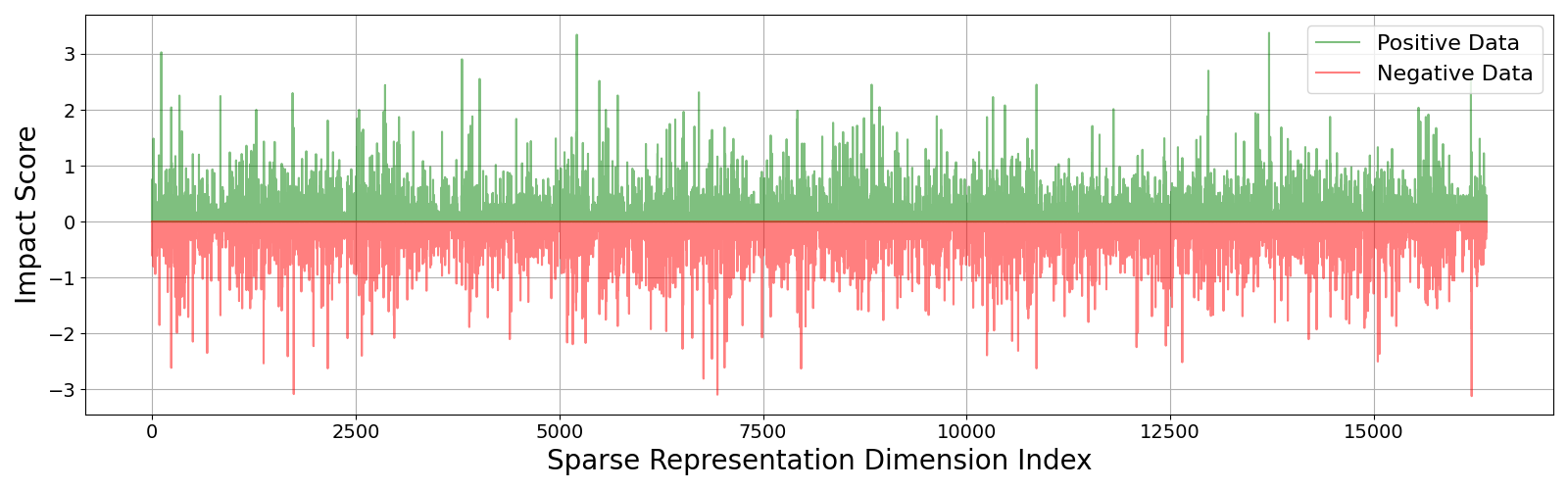}
    \caption{
    Scores of the positive and negative impacts of each sparse representation dimension in \textbf{truthfulness} domain.
 }
    \label{fig:dimension_truthfulness}
    % \vspace{-2mm}
\end{figure*}

We visualized the impact scores of each sparse feature dimension on different domains, computed from the positive and negative data in Eq.~\ref{eq:kl} in the sparse space. The results for the safety domain are shown in Fig.~\ref{fig:dimension_safety}, for the fairness domain in Fig.~\ref{fig:dimension_fairness}, and for the truthfulness domain in Fig.~\ref{fig:dimension_truthfulness}.

\subsection{Impact of Hyper-parameter $\alpha$}\label{sec:effect_scale}

We evaluate the steering effect under different $\alpha$. Score represents the attribute score, measuring how well the model output aligns with the targeted property (see Sec.~\ref{sec:metrics} for details). Coherence quantifies the semantic consistency between the steered output and the user input, since excessively large steering strength 
$\alpha$ may distort the generated content. To jointly capture both aspects, we further considered the product of Score and Coherence, which balances property alignment and semantic preservation under different $\alpha$ values.

\begin{figure*}[t]
    \centering
    % \vspace{-2mm}
    \includegraphics[width=1\textwidth]{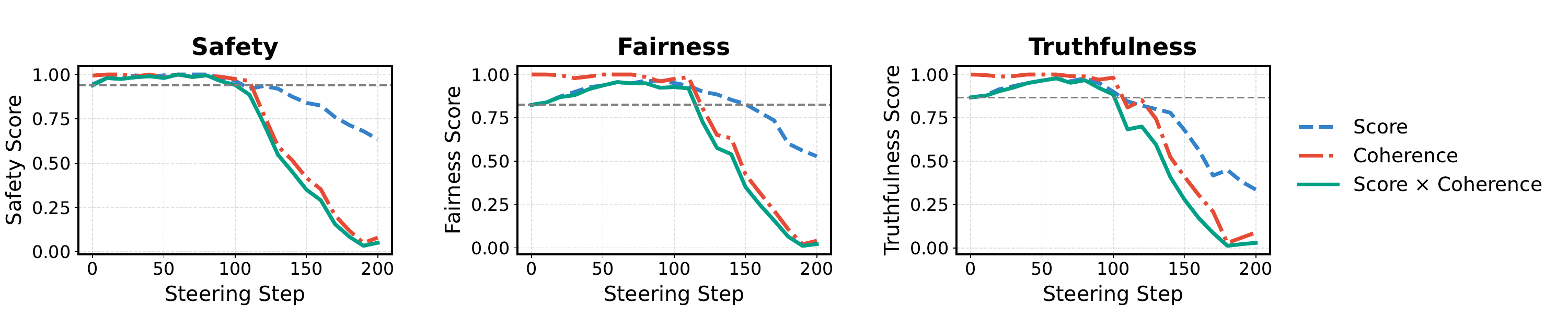}
    \caption{Effect of \shortname~in three domains with steering scale ranging from 0 to 200 with step=10 on Gemma-2B-it.
 }
    \label{fig:effect_scale}
    % \vspace{-2mm}
\end{figure*}

The results  are demonstrated in Fig.~\ref{fig:effect_scale}. As $\alpha$ increases, the attribute Score steadily improves. For instance, in the safety domain, the Score rises from approximately 0.94 at $\alpha=10$ to about 0.96 at 
$\alpha=80$ , indicating enhanced property control. However, Coherence exhibits a decline when $\alpha$ becomes too large: while Coherence remains above 0.95 for moderate values (
$\alpha \leq 50$), it drops to around 0.90 once 
($\alpha \geq 90$), suggesting that strong interventions compromise semantic fidelity. By examining the combined metric Score*Coherence, we identify an optimal trade-off: the product reaches its maximum around
$\alpha=60$, where the model achieves both high property scores and stable coherence. Overall, we set $\alpha=60$ for all tasks by default.

\subsection{Explanations for Top-K Identified Features}\label{appendix:neuronpedia}

Neuronpedia explanation results on fairness and truthfulness domains are shown in Tab.~\ref{tab:neuronpedia_fairness} and Tab.~\ref{tab:neuronpedia_truthfulness}, respectively.
As shown in Tab.~\ref{tab:neuronpedia_fairness}, the $K_+$ features primarily correspond to semantics related to social harmony, positive emotions, and respectful or inclusive expressions, reflecting contexts that promote fairness and equality.  
In contrast, the $K_-$ features are dominated by concepts associated with negative judgment, discrimination, or hostile descriptions of individuals or groups, which are negatively correlated with fair and unbiased outputs.

As shown in Tab.~\ref{tab:neuronpedia_truthfulness}, the sparse features with high positive weights ($K_+$) tend to capture linguistic patterns related to epistemic uncertainty, logical conditions, and causal reasoning, such as references to “knowledge,” “existence,” and “if-then” structures. These features generally correlate with cautious, truth, seeking language that avoids overstatement.  
Conversely, the negatively weighted features ($K_-$) are more aligned with formulaic or technical language, including statistical expressions, programming, related terminology, or document structural markers. These may reflect abstract or rigid content that lacks clear grounding in factual assertions, thereby being less associated with truthful or sincere expression in conversational contexts.

\begin{table*}[tb]
\centering
\setlength{\tabcolsep}{4pt} % 调整列间距
\renewcommand{\arraystretch}{1.1} % 调整行高
\small % 使用较小字号
\begin{tabular}{@{}c c c p{0.71\textwidth} S[table-format=2.2]@{}}
\toprule
\textbf{Group} & \textbf{Rank} & \textbf{Index} & \textbf{Explanation of SAE Feature} & \textbf{Weight} \\
\midrule
\multirow{7}{*}{$K_+$} 

& 1  & 6923 &elements related to social dynamics and interpersonal relationships  & 6.46 \\
& 2 & 6784 & expressions of happiness and celebration & 5.38 \\
& 3 & 4291 &conditional or situational phrases in contexts that may imply restrictions or guidelines & 3.51 \\
& 4 & 14216& phrases that indicate classifications or types with a focus on personal experiences& 3.26 \\
& 5 & 1968 &positive descriptors and evaluations of people or things &3.11 \\
& 6 &5074&features related to medical terminology and health conditions &2.96\\
& 7 &4781 &  references to emotional states and interpersonal relationships&2.85 \\
\midrule
\multirow{7}{*}{$K_-$} 

& 1  & 1218& negative descriptions and issues related to outcomes and performance                      & -5.45 \\
& 2 & 10549 &  negative sentiments or harmful concepts in various contexts& -5.30 \\
& 3 &  12051 & negative descriptions or reviews of experiences& -4.65 \\
& 4 &  7710&incidents of crime and violence depicted in a societal context & -4.41 \\
& 5 & 9534 &  concepts related to negative outcomes and their implications& -3.87 \\
& 6 & 10623&  topics related to societal judgment and stigma, particularly concerning women and parenting& -3.81 \\
& 7 &1495 & negative descriptors and insults directed toward individuals or groups& -3.67 \\

\bottomrule
\end{tabular}
\caption{Top sparse features identified by \shortname~for the \textbf{fairness} domain on \texttt{Gemma-2-2B-it}. 
Feature interpretations are obtained from Neuronpedia~\cite{neuronpedia}, and the corresponding weights are learned by \shortname.}
\label{tab:neuronpedia_fairness}
\end{table*}

\begin{table*}[tb]
\centering
\setlength{\tabcolsep}{4pt} % 调整列间距
\renewcommand{\arraystretch}{1.1} % 调整行高
\small % 使用较小字号
\begin{tabular}{@{}c c c p{0.71\textwidth} S[table-format=2.2]@{}}
\toprule
\textbf{Group} & \textbf{Rank} & \textbf{Index} & \textbf{Explanation of SAE Feature} & \textbf{Weight} \\
\midrule
\multirow{7}{*}{$K_+$} 

& 1  & 13713 & inquiries about knowledge and uncertainty regarding events or situations  & 3.37 \\
& 2 & 5215 &  discussions of legal and social concepts related to guilt and innocence
& 3.34 \\
& 3 & 114 &  references to the presence or absence of specific entities or conditions in various contexts& 3.02 \\
& 4 &  3805&conditional statements checking for variable existence or conditions
& 2.90 \\
& 5 & 12968&conjunctions and transition words indicative of causal relationships&2.69 \\
& 6 &4022&references to events or actions related to conflict, struggle, or disruption in narrative contexts&2.55\\
& 7 &16191 &digital and numerical data representations
 &2.52 \\
\midrule
\multirow{7}{*}{$K_-$} 

& 1  & 16200 & terms and concepts related to statistical methods and analysis                      & -3.12\\
& 2 & 6941 & terms related to scientific studies and methodologies
 & -3.09 \\
& 3 & 1740&references to data processing and analysis methodologies
& -3.08 \\
& 4 & 6770 & questions and discussions regarding product features and their implications
& -2.81 \\
& 5 & 7968 & the beginning of sections and titles in structured documents
& -2.63 \\
& 6 &10858&  phrases related to safety measures and inventions designed to prevent accidents& -2.62 \\
& 7 & 2157 &  keywords and identifiers related to programming and software development concepts& -2.61 \\

\bottomrule
\end{tabular}
\caption{Top sparse features identified by \shortname~for the \textbf{truthfulness} domain on \texttt{Gemma-2-2B-it}. 
Feature interpretations are obtained from Neuronpedia~\cite{neuronpedia}, and the corresponding weights are learned by \shortname.}
\label{tab:neuronpedia_truthfulness}
\end{table*}

%%%%%%%%%%%%%%%%%%%%%%%%%%%%%%%%%%%%%%%%%%%%%%%%%%%%%%%%%%%%%%%%%%%%%%%%%%%%%%%%
\end{sloppypar}
\end{document}